\documentclass[twocolumn,tighten]{aastex62}

\usepackage{graphicx}
\usepackage{grffile}
\usepackage{natbib}
\usepackage{amsmath}
\usepackage{url}
\usepackage{enumitem}
\usepackage{tabularx}
\usepackage{multirow}
\newcommand{\wm}{W\,m$^{-2}$\;}
\newcommand{\wmn}{W\,m$^{-2}$}
\newcommand{\gkg}{g\,kg$^{-1}$\;}
\newcommand{\gkgn}{g\,kg$^{-1}$}
\newcommand{\so}{S$_{0}$\;}
\newcommand{\son}{S$_{0}$}

\renewcommand{\nat}{Natur}

\newcommand{\natco}{NatCo}

\newcommand{\jas}{JAtS}
\newcommand{\ana}{AN}
\newcommand{\asbio}{AsBio}
\renewcommand{\icarus}{Icar}
\renewcommand{\apj}{ApJ}
\renewcommand{\apjl}{ApJL}
\renewcommand{\jgr}{JGR}
\newcommand{\jgrd}{JGRD}
\newcommand{\cldy}{ClDy}
\renewcommand{\grl}{GeoRL}

\newcommand{\ato}{AtO}
\newcommand{\jcli}{JCli}

\newcommand{\five}{NG5~}
\newcommand{\nine}{NG9~}

\accepted{for publication in Icarus}

\begin{document}
	
	\title{Climate Sensitivity to Ozone and its relevance on the Habitability of Earth-like planets}
	\shorttitle{Ozone and Habitability}
	\shortauthors{I.\,Gomez-Leal et al.}
	\author{Illeana Gomez-Leal}
	\affiliation{Cornell Center for Astrophysics \& Planetary Science, 304 Space Sciences Bldg., Cornell University Ithaca, NY14853-6801, USA}
	\affiliation{Carl Sagan Institute, 304 Space Sciences Bldg., Cornell University, Ithaca, NY14853-6801, USA}
	\author{Lisa Kaltenegger}
	\affiliation{Cornell Center for Astrophysics \& Planetary Science, 304 Space Sciences Bldg., Cornell University Ithaca, NY14853-6801, USA}
	\affiliation{Carl Sagan Institute, 304 Space Sciences Bldg., Cornell University, Ithaca, NY14853-6801, USA}
	\author{Valerio Lucarini}
	\affiliation{Department of Mathematics and Statistic, Whiteknights, PO Box 220, University of Reading, Reading RG6 6AX, UK}
	\author{Frank Lunkeit}
	\affiliation{CEN, Institute of Meteorology, University of Hamburg Grindelberg 5, Hamburg D-20144, Germany}
	
	\email{illeana@astro.cornell.edu; gomezleal.gaia@gmail.com}
	
	\begin{abstract}
		\makeatletter{}Atmospheric ozone plays an important role on the temperature structure of the atmosphere. However, it has not been included in previous studies on the effect of an increasing solar radiation on the Earth’s climate. Here we study the climate sensitivity to the presence/absence of ozone with an increasing solar forcing for the first time with a global climate model. We show that the warming effect of ozone increases both the humidity of the lower atmosphere and the surface temperature. Under the same solar irradiance, the mean surface temperature is 7 K higher than in an analog planet without ozone. Therefore, the moist greenhouse threshold, the state at which water vapor becomes abundant in the stratosphere, is reached at a lower solar irradiance (1572~\wm with respect to 1647~\wm in the case without ozone). Our results imply that ozone reduces the maximum solar irradiance at which Earth-like planets would remain habitable.

	\end{abstract}
	
	
	\NewPageAfterKeywords

\section{Introduction}
\label{sec:introduction}
\makeatletter{}The intensification of the solar luminosity with time will increase Earth surface temperatures and may cause the loss of the planet's water, threatening its habitability \citep[e.g.][]{Kasting1984, Kasting1988, Kasting1993, Kopparapu2013}. It is still an open question which process will dominate: the moist greenhouse effect or the runaway greenhouse effect. Water vapor is scarce in Earth stratosphere at the present solar irradiance. However, at higher temperatures, if the troposphere is near the saturation level, water vapor might considerably increase in the stratosphere at a certain level of solar irradiance \citep{Ingersoll1969}. \citet{Kasting1993} identifies this moist greenhouse threshold (MGT) with a dramatic increase in the stratospheric water vapor mixing ratio with the solar forcing. Then, water vapor is photodissociated in the stratosphere by solar UV radiation, hydrogen escapes, and water gradually disappears from the planet \citep{Towe1981}. In a runaway greenhouse state, the evaporation of the oceans leads to a steamy atmosphere, surface temperature rises above 1800~K \citep[e.g.][]{Kopparapu2013}, destabilizing the climate, and the water contain of the planet is rapidly lost. The moist greenhouse state and the runaway greenhouse state are used to define the inner boundary of the conservative Habitable Zone \citep[e.g.][]{Kasting1988, Kasting1993, Kopparapu2013, Ramirez2014, Ramirez2016}. The characterization of the radiative conditions that lead to these states is crucial to understand the evolution of our planet and the habitability conditions of exoplanets \citep[e.g.][]{Abe2011, Words2013, Yang2014}. 

\cite{Kasting1993} using a radiative-convective model attained the MGT at a solar irradiance of about 1500~\wm (1.1~\son\footnote{\son=1361~\wm is the solar constant.}) and a water mixing ratio of 3~\gkg in the stratosphere. However, GCM results on the habitability of Earth-like planets by different models diverge. According to \cite{Leconte2013} (hereafter L13), Earth-like planets with an atmosphere composed by 1 bar of N$_{2}$, 376~ppm of CO$_{2}$, and a variable amount of water vapor do not attain a moist greenhouse state, and the runaway greenhouse effect occurs at about 1500~\wm (which corresponds to a distance of 0.95~au in the present in the Solar System). However, \cite{Wolf2015} (hereafter W13) finds two possible moist greenhouse states: one at 1531~\wm (0.94~au), coinciding with a maximum of the climate sensitivity and another at about 1620~\wm (0.92~au) following the water vapor mixing ratio of 3~\gkg given by \cite{Kasting1993}. These two models also show a large surface temperature difference at the present solar irradiance for a planet without O$_{3}$. L13 simulations have a surface temperature of 283~K, while W13 shows a temperature of 289~K, which is similar to the present Earth's mean value. In addition, simulations increasing the solar forcing on an Earth-like planet using either 1D \citep{Kopparapu2013, Kasting2015} or 3D models \citep{Leconte2013, Wolf2015, Wolf2017} do not include ozone in the atmosphere, and therefore, both the temperature and the water mixing ratio at their initial state differ greatly from those of present Earth. Aquaplanet simulations including ozone have shown a better agreement with Earth values \citep{Popp2016}, but they did not include sea ice and continents, which have a significant effect on the albedo and the temperature of the planet.\\

Ozone is a greenhouse gas and plays a key role in the energy balance of the Earth. It absorbs most of the solar UV radiation through photodissociation, protecting surface life from genetic damage and warming the stratosphere. The resultant temperature structure determines the level of the tropopause \citep[e.g.][]{Wilcox2012}. In the last decades, it has contributed to the radiative forcing of our planet with about 0.35~\wmn, due to its increase in the troposphere by human activities \citep{Forster2007}, and about -0.05~\wmn, due to its decrease in the stratosphere. By comparison, it is 20\% the radiative forcing induced by carbon dioxide. The concentration change produced by the 11-year solar cycle generates a small radiative forcing (0.004~\wmn) \citep{Gray2009}, but the contribution of atmospheric ozone might become more important at a larger solar forcings. The rise in water vapor has chemical and radiative effects on the atmosphere. The products of its dissociation, such HO$_{x}$ radicals, increase, depleting ozone concentration. However, they also remove NO$_{2}$ by increasing HNO$_{3}$, which slows O$_{3}$ depletion. At the same time, water vapor absorbs latent heat, cooling the environment and decreasing the reaction rates. Several studies have proved that, by the combination of these effects, increasing water vapor in the atmosphere at the same solar irradiance, only depletes ozone in the tropical lower stratosphere and the high latitudes of the southern hemisphere, while elsewhere ozone increases \citep[e.g.][]{Evans1998, Tian2009}. \citet{Bordi2012} showed that the removal of ozone from Earth's atmosphere induces considerable changes on climate: the structure of the stratosphere is modified, the stratification disappears, convection reaches higher altitudes, and the tropopause level rises. These effects cool the planet and the water vapor content in the atmosphere decreases.

Ozone is part of the atmospheric composition of other planets in the Solar System \citep[e.g.][]{Lane1973, Fast2006, Montmessin2011} and although the origin of Earth's ozone layer is biotic \citep[e.g.][]{Kasting1980, Kasting1981}, O$_{3}$ can also be produced abiotically \citep[e.g.][]{Canuto1983, Finney2016} and it could possibly be build up in atmospheres with a reduced chemical composition under conditions of high ultraviolet radiation \citep[e.g.][]{Domagal2010, Domagal2014}. Therefore, ozone might have relevant implications on the greenhouse effect and the planetary habitability of terrestrial planets.\\

Here we study the climate sensitivity to the increase of solar irradiance by the comparison of the radiative effect of the actual atmospheric concentration of ozone with an equivalent atmosphere without ozone. In Section~\ref{conv}, we briefly present the problem of cumulus parameterization in global climate models. In Section~\ref{mod}, we describe the model and the methods used. In Section~\ref{res}, we present our results. First, we study the climate sensitivity of the present Earth with and without ozone and then, we increase the solar irradiance in subsequent simulations until the atmosphere becomes opaque. We study the evolution of temperature, humidity, and cloud formation in these two scenarios. For the first time with a 3D model, we determine the MGT by measuring the increase of the water mixing ratio in the low stratosphere including atmospheric ozone. Finally, we compare our results with Earth reanalysis data, with other types of cumulus parameterization, and with previous GCM studies. We show that ozone increases the temperature and the humidity of the stratosphere and the troposphere, and as a consequence, the MGT is reached at lower solar irradiance (and larger distance) when the radiative effect of ozone is included in the model. Therefore, the atmospheric ozone concentration may have an important role on the habitability of Earth-like planets.

 \section{Cumulus convection parameterization}\label{conv}

The scale of the physical processes involved in moist convection range from a few kilometers to some microns, being smaller than the spatial resolution of any climate model. For this reason, these processes are parameterized. Convective parameterization schemes compute the cloud formation in a model column, containing convective clouds of varying size and height. The atmospheric variables are separated as $\varepsilon=\overline{\varepsilon}+\varepsilon'$, where $\overline{\varepsilon}$ is the spatial average over the large region and $\varepsilon'$ represents the variable in the cloud scale. The equations for the dry static energy\footnote{The dry static energy is defined as $s\equiv c_{p}T+gz$, where $c_{p}$ is the specific heat at a constant pressure, $T$ is the temperature, $g$ is the gravitational acceleration, and $z$ is the altitude.} ($s$) and the water vapor content ($q$) are:

\begin{equation} 
\begin{cases} 
\frac{\partial \overline{s}}{\partial t}+\vec{v}\cdot{\nabla s}+ \overline{\omega}~\frac{\partial \overline{s}}{\partial p}= - \frac{\partial \overline{\omega's'}}{\partial p}+  L(\overline{C}-\overline{E})+ \overline{Q}_{R}\\
\frac{\partial \overline{q}}{\partial t}+\vec{v}\cdot{\nabla q}+ \overline{\omega}~\frac{\partial \overline{q}}{\partial p}= - \frac{\partial \overline{\omega'q'}}{\partial p}-(\overline{C}-\overline{E})
\end{cases}
\end{equation}

where $\vec{v}$ is the horizontal velocity field, $\omega$ is the vertical velocity in the pressure coordinate $p$, $L$ is the latent heat, $Q_{R}$ is the radiative forcing, $C$ is the condensation rate, and $E$ is the evaporation rate. The schemes estimate the triggering of convection, as well as the vertical structure and the magnitude of $s$ and $q$, representing the total convective activity. They are classified into three families: i) adjustment schemes \citep[e.g.][]{Manabe1965, Betts1986, Frierson2007}, which are based on the idea that convection acts in order to adjust the state to a reference profile that is usually prescribed or computed to match observations; ii) moisture convergence schemes \citep[e.g.][]{Kuo1965, Kuo1974}, which are based on the idea that convection acts in order to store a certain fraction of moisture $\beta$ and to precipitate the remaining (1-$\beta$); and iii) mass-flux schemes \citep[e.g.][]{Arakawa1974, Kain2004}, more complex than the two precedent parameterization types, relate heat and moisture to cloud physical processes. The collective behavior of cumulus clouds in each air column is represented by a bulk cloud. The mass-flux of the cloud is represented by the amount of air transported in the vertical direction inside the cloud, the entrainment and the detrainment rates of environmental air into and out of the cloud, respectively, and it can be extended to add other cloud dynamics such as updrafts and downdrafts \citep[e.g.][]{Tiedtke1989}. 

These three types of parameterization give similar results for Earth's climate \citep[e.g.][]{Tiedtke1988, Arakawa2004}. However, the intrinsic differences of their structure, the non-resolved processes involved in cumulus convection, and the uncertainties related to the climate evolution may produce different results in extreme conditions. L13 and W15 clearly differ on the climate of an Earth-like planet without ozone a higher solar radiations, as we point above. Among other differences, L13 uses an adjustment scheme and W15 a mass flux scheme. A further effort should be made in order to understand the possible bias of these parameterizations and their influence on climate. In Section~\ref{res}, we compare the results of these two studies with Earth reanalysis data, we analyse PlaSim simulations using an adjustment scheme and a moisture convergence scheme, and we compare them with the simulations in W15.

 %%%%%%%%%%%%%%%%%%%%
 %TAB1
 %%%%%%%%%%%%%%%%%%%%
 \begin{table*}
 	\begin{center}
 		\centering
 		\begin{tabularx}{1.\textwidth}{c  c l c   c   c  l c  c   c   c  l c   c   }
 			\multicolumn{11}{l}{\hspace{1.5cm}$\emph{The present Earth's climate}$}\\\cline{2-11}
 			&\multicolumn{1}{|p{1.5cm}}{\centering Model} &\multicolumn{1}{|p{1cm}}{\centering $O_{3}$} & \multicolumn{1}{p{1.8cm}}{\centering $TSI$(Wm$^{-2}$)} &\multicolumn{1}{p{2cm}}{\centering [$CO_{2}$](ppm)} &\multicolumn{1}{|p{1cm}}{\centering $T_{S}$(K)}  &\multicolumn{1}{p{1.cm}}{\centering $T_{eff}$(K)} & \multicolumn{1}{p{1cm}}{\centering A} & \multicolumn{1}{p{0.5cm}}{\centering $g_{n}$}& \multicolumn{1}{|p{1cm}}{\centering T$_{40}$(K)} &\multicolumn{1}{p{2cm}|}{\centering $q_{r}$(\gkgn)} \\\cline{1-11}
 			
 			\multicolumn{1}{|c}{\centering }&\multicolumn{1}{|c}{ERA}  	      & \multicolumn{1}{|c}{yes}& \multicolumn{1}{c}{1361} & 388  & \multicolumn{1}{|c}{289.1}  &\multicolumn{1}{c}{255.3}  & 0.294 & 0.392  & \multicolumn{1}{|c}{216} & \multicolumn{1}{c|}{2.3$\times$10$^{-3}$}\\\cline{1-11}
 			\multicolumn{1}{|c}{\centering a}&\multicolumn{1}{|c}{PlaSim} 		& \multicolumn{1}{|c}{yes} & \multicolumn{1}{c}{1361}   & 388  & \multicolumn{1}{|c}{291.0} & \multicolumn{1}{c}{255.2} & 0.296 & 0.419 & \multicolumn{1}{|c}{217} & \multicolumn{1}{c|}{7.4$\times$10$^{-3}$}  \\
 			\multicolumn{1}{|c}{\centering b}&\multicolumn{1}{|c}{PlaSim}  		& \multicolumn{1}{|c}{no} & \multicolumn{1}{c}{1361}  & 388   & \multicolumn{1}{|c}{284.0} & \multicolumn{1}{c}{251.8} & 0.334 & 0.394 & \multicolumn{1}{|c}{180} & \multicolumn{1}{c|}{5.0$\times$10$^{-4}$}\\\cline{1-11}
 			\multicolumn{1}{|c}{\centering c}&\multicolumn{1}{|c}{PlaSim}  		& \multicolumn{1}{|c}{yes}& \multicolumn{1}{c}{1361}& 280  & \multicolumn{1}{|c}{290.3} & \multicolumn{1}{c}{254.8} & 0.301  & 0.406& \multicolumn{1}{|c}{220} & \multicolumn{1}{c|}{4.4$\times$10$^{-3}$}\\
 			\multicolumn{1}{|c}{\centering d}&\multicolumn{1}{|c}{PlaSim}  		& \multicolumn{1}{|c}{yes}& \multicolumn{1}{c}{1361}& 560  & \multicolumn{1}{|c}{293.3} & \multicolumn{1}{c}{255.7} & 0.288  & 0.430& \multicolumn{1}{|c}{215} & \multicolumn{1}{c|}{13$\times$10$^{-3}$}\\\cline{1-11}
 			
 			\multicolumn{11}{l}{\hspace{1.5cm}$\emph{The Moist Greenhouse Threshold}$}\\\cline{2-11}
 			&\multicolumn{1}{|p{1.5cm}}{\centering Model} &\multicolumn{1}{|p{1cm}}{\centering $O_{3}$} & \multicolumn{1}{p{1.8cm}}{\centering $TSI$(Wm$^{-2}$)} &\multicolumn{1}{p{2cm}}{\centering [$CO_{2}$](ppm)} &\multicolumn{1}{|p{1cm}}{\centering $T_{S}$(K)}  &\multicolumn{1}{p{1.cm}}{\centering $T_{eff}$(K)} & \multicolumn{1}{p{1cm}}{\centering A} & \multicolumn{1}{p{0.5cm}}{\centering $g_{n}$} & \multicolumn{1}{|p{1cm}}{\centering T$_{st}$(K)} &\multicolumn{1}{p{2cm}|}{\centering $q_{r}$(\gkgn)} \\\cline{1-11}
 			\multicolumn{1}{|c}{\centering e}&\multicolumn{1}{|c}{PlaSim} & \multicolumn{1}{|c}{yes}& \multicolumn{1}{c}{1572} & 388  & \multicolumn{1}{|c}{319.9}  &\multicolumn{1}{c}{269.5} & 0.238 & 0.496 &\multicolumn{1}{|c}{243} &\multicolumn{1}{c|}{7.8}\\
 			\multicolumn{1}{|c}{\centering f}&\multicolumn{1}{|c}{PlaSim}  & \multicolumn{1}{|c}{no}& \multicolumn{1}{c}{1572} & 388  & \multicolumn{1}{|c}{311.8}  &\multicolumn{1}{c}{266.1} & 0.277 & 0.470 &\multicolumn{1}{|c}{224} &\multicolumn{1}{c|}{0.6}\\\cline{1-11}
 			\multicolumn{1}{|c}{\centering g}&\multicolumn{1}{|c}{PlaSim} & \multicolumn{1}{|c}{no}& \multicolumn{1}{c}{1647} & 388 & \multicolumn{1}{|c}{319.8}  &\multicolumn{1}{c}{271.0} & 0.255 & 0.484 &\multicolumn{1}{|c}{240} &\multicolumn{1}{c|}{6.8}\\\cline{1-11}
 			
 		\end{tabularx}
 		
 		\caption{Comparison of the Earth's present state and the Moist Greenhouse threshold mean global data. The ozone ($O_{3}$) concentration, the total solar irradiance ($TSI$), and the $CO_{2}$ concentration (in ppm) are initial conditions. The surface temperature ($T_{S}$), the effective temperature ($T_{eff}$), the Bond albedo (A), the normalized greenhouse parameter ($g_{n}$) calculations are explained in Section~\ref{mod}. The temperature (T$_{40}$) and the water vapor mixing ratio ($q_{r}$) are both measured at a pressure level of 40h~Pa.}\label{tab1}
 	\end{center}
 \end{table*} 
 %%%%%%%%%%%%%%%%%
 
 \section{Methods}\label{mod}
 We have used the intermediate complexity model Planet Simulator (PlaSim)\citep{Fraedrich2005, Lunkeit2011}\footnote{PlaSim is freely available at \url{https://www.mi.uni-hamburg.de/en/arbeitsgruppen/theoretische-meteorologie/modelle/plasim.html}} to simulate the global warming of the Earth under an increasing solar irradiance. While being simpler than the state-of-the-art climate models in terms of resolution and adopted parameterizations, this type of models represent a compromise between complexity and calculation time. They can simulate a large variety of scenarios and allow us to examine certain aspects of the climate in a very efficient manner, performing a large number of simulations in a short time. As a result, the model has been instrumental in studying climate change using rigorous methods of statistical mechanics \citep{Ragone2016}.  It has the advantage of featuring a great degree of flexibility and robustness when terrestrial, astronomical, and astrophysical parameters are altered. It has been extensively used for studying paleoclimatic conditions, exotic climates, and circulation regimes potentially relevant for exoplanets \citep{Lucarini2013, Boschi2013, Linsen2015}.
 
 The primitive equations for vorticity, divergence, temperature, and surface pressure are solved via the spectral transform method \citep{Eliasen1970, Orszag1970}. The parameterization in the shortwave (SW) radiation uses the ideas of \cite{Lacis1974} for the cloud free atmosphere. Transmissivities and albedos for high, middle, and low level clouds are parameterized following \cite{Stephens1978} and \cite{Stephens1984}. The downward radiation flux density $F^{\downarrow SW}$ is the product of different transmission factors with the solar flux density ($\mathcal{E}_{0}$) and the cosine of the solar zenith angle ($\mu_{0}$) as
 \begin{equation}
 F^{\downarrow SW}=\mu _{0}\mathcal{E}_{0}\cdot \mathcal{T}_{R}\cdot \mathcal{T}_{H_{2}O}\cdot \mathcal{T}_{O_{3}}\cdot \mathcal{T}_{C}\cdot \mathcal{R}_{S}
 \end{equation}
 which includes the transmissivities due to Rayleigh scattering ($R$), and cloud droplets ($C$), and $R_{S}$ comprises different surface albedo values. $\mathcal{E}_{0}$ and $\mu_{0}$ are computed following \cite{Berger1978a, Berger1978b}. For the clear sky longwave (LW) radiation (F$^{LW}$), the broad band emissivity method is employed \citep{Manabe1961, Rodgers1967, Sasamori1968, Katayama1972, Boer1984},
 \begin{equation}
 F^{\uparrow LW}(z)= \mathcal{A}_{S}B(T_{S}) \mathcal{T}_{(z,0)}+\int_0^z B(T')\frac{\delta \mathcal{T}_{(z,z')}}{\delta z'}
 \end{equation}
 \begin{equation}
 F^{\downarrow LW}(z)= \int_\infty^z B(T')\frac{\delta \mathcal{T}_{(z,z')}}{\delta z'}
 \end{equation}
 where $B(T)$ denotes the blackbody flux and $\mathcal{A_{S}}$ is the surface emissivity. The transmissivities for water vapor, carbon dioxide, and ozone are taken from \cite{Sasamori1968}. The empirical formulas are obtained from meteorological data and are dependent on the effective amount of each gas. The effective amount is obtained as
 \begin{equation}
 u_{X}(p,p')=\frac{1}{g}\int_{p}^{p'}q_{X}(\frac{p''}{p_{0}})dp''
 \end{equation}
 where $g$ is the gravitational acceleration, $q_{X}$ is the mixing ratio, $p$ is the pressure, $p_{0}=1000~hPa$ is the reference pressure.
 
 The H$_{2}$O continuum absorption is parameterized as
 \begin{equation}
 \tau _{cont}^{H_{2}O}=1.- exp(-0.03 u_{H_{2}O})
 \end{equation}
 
 To account for the overlap between the water vapor and the carbon dioxide bands near 15~\micron, the CO$_{2}$ absorption is corrected by a H$_{2}$O transmission at 15~\micron  given by
 \begin{equation}
 \mathcal{T}^{15\micron}_{H_{2}O}=1.33 - 0.832(u_{H_{2}O}+ 0.0286)^{0.26}
 \end{equation}
 
 Cloud flux emissivities are obtained from the cloud liquid water content \citep{Stephens1984} by
 \begin{equation}
 \mathcal{A}^{cl}=1.- exp(-\beta_{d}k^{cl}W_{L})
 \end{equation}
 where $\beta_{d}=1.66$ is the diffusivity factor, $k^{cl}$ is the mass absorption coefficient, set to a default value of 0.1 m$^{2}$g$^{-1}$  \citep{Slingo1991}, and $W_{L}$ is the cloud liquid water path. For a single layer between z and z' with the fractional cloud cover $\zeta$, the total transmissivity is
 \begin{equation}
 \mathcal{T}^{*}_{(z,z')}=\mathcal{T}_{(z,z')}(1.- \zeta \mathcal{A}^{cl})
 \end{equation}
 where $\mathcal{T}_{(z,z')}$ is the clear sky transmissivity. Random overlapping of clouds is assumed for multilayers and the total transmissivity becomes
 \begin{equation}
 \mathcal{T}^{*}_{(z,z')}=\mathcal{T}_{(z,z')}\Pi_{j}(1.- \zeta_{j} \mathcal{A}^{cl}_{j})
 \end{equation}
 where $j$ denotes each cloud layer.
 
 It includes dry convection, large-scale precipitation, boundary-layer fluxes of latent and sensible heat, and vertical and horizontal diffusion \citep{Louis1979, Laursen1989, Roeckner1992}. Penetrative cumulus convection is simulated by a moist convergence scheme \citep{Kuo1965, Kuo1974} including some improvements: cumulus clouds are assumed to exist only if the environmental air temperature and moisture are unstable stratified with respect to the rising cloud parcel, and the net ascension is positive. Shallow convection is represented following \cite{Tiedtke1988} and clouds originated by extratropical fronts are simulated considering the moisture contribution between the lifting level and the top of the cloud, instead of the total column. Cumulus convection can be simulated using a Betts-Miller adjustment scheme \citep{Betts1986, Frierson2007} instead (see Section~\ref{res} for a comparison of the results using Kuo and Betts-Miller schemes). \\

 \begin{figure*}[ht]\centering  
 	\includegraphics[width=0.9\textwidth]{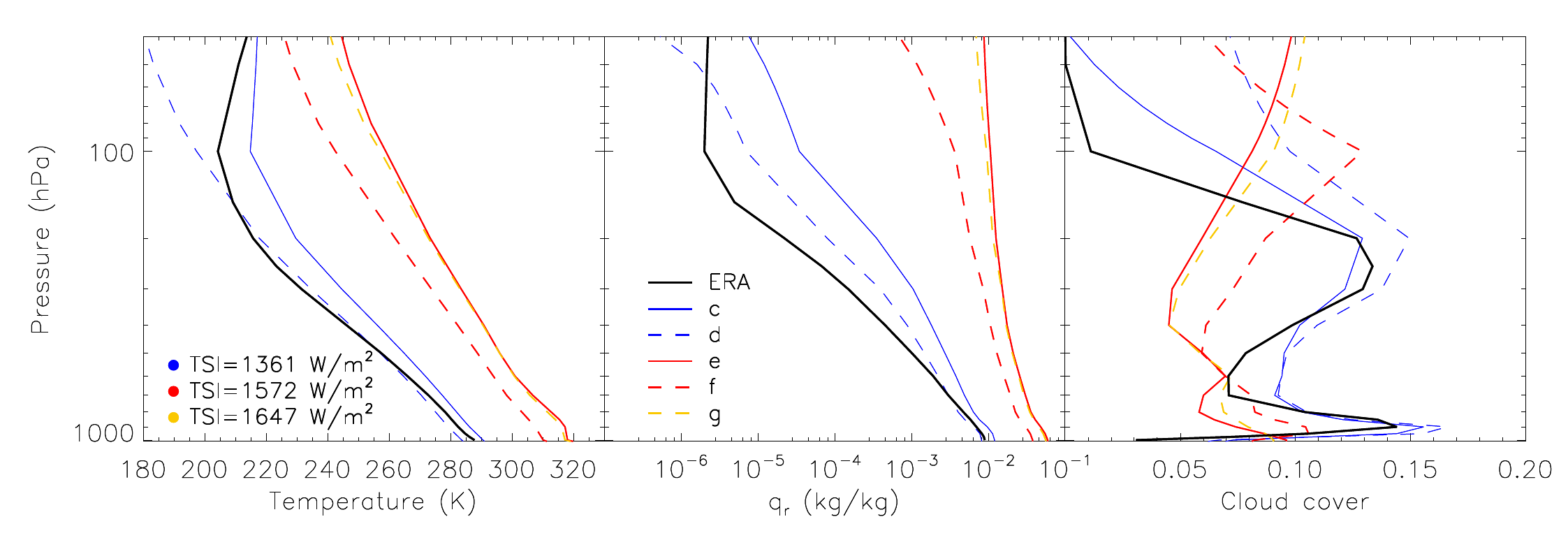}
 	\caption{Mean atmospheric profiles with and without $O_{3}$ at several total solar irradiance ($TSI$) values. Mean temperature (left), mean water vapor mixing ratio $q_{r}$ (middle), and mean cloud cover (right) for PlaSim simulations with ozone (solid coloured lines) and without ozone (dashed lines) in comparison with ERA (solid black). The simulations \emph{c}, \emph{d}, \emph{e}, \emph{f}, and \emph{g} correspond to those listed in Table~\ref{tab1}. The moist greenhouse threshold is attained at 1572~\wm in the presence of ozone (solid red) and at 1647 Wm$^{-2}$ in the absence of ozone (dashed yellow). \label{prof_era}}
 \end{figure*}
 
 \begin{figure}[hb]
 	\includegraphics[width=1.\columnwidth]{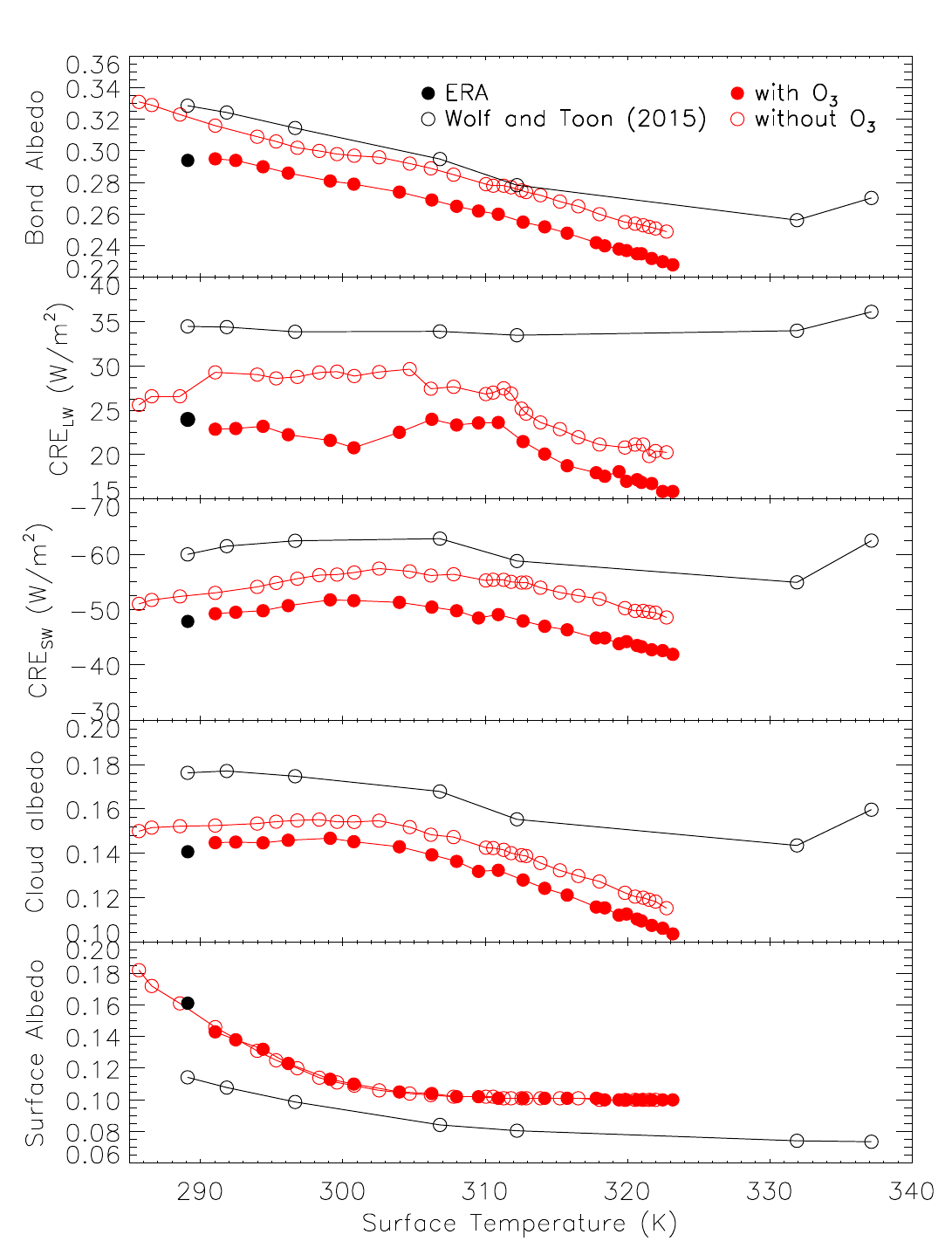}     
 	\caption{Comparison between our simulations, ERA data, and the results in \citet{Wolf2015}. From top to bottom: Bond albedo, cloud radiative effect ($CRE$) for the longwave (LW) radiation, $CRE$ for the shortwave (SW) radiation, cloud albedo, and surface albedo vs. surface temperature. \label{cre_era}} 
 \end{figure}

The model includes a 50 m mixed-layer ocean and a thermodynamic sea-ice model. The effects of water, carbon dioxide, and ozone are taken in account in the radiative transfer. The ozone concentration is prescribed following the distribution described by \citet{Green1964},
 \begin{equation}\label{eq11}
 u_{O_{3}}(z)=(a+a e^{-b/c})/(1+e^{(z-b)/c})
 \end{equation}
 where $u_{O_{3}}(z)$ is the ozone concentration in a vertical column above the altitude $z$, $a$ is the total ozone in the vertical column above the ground, and $b$ the altitude where the ozone concentration is maximal, and $c$ is a fitting parameter. Equation~\ref{eq11} fits to the midlatitude winter ozone distribution with $a= 0.4$~cm, $b= 20$~km, and $c= 5$~km. The latitudinal variation and the annual cycle are modeled by, 
 \begin{equation}
 a(t,\phi)=a_{0}+a_{1}\,|sin\phi|+a_{c}\,|sin\phi|\,cos[(2\pi/n)(d-d_{off})] 
 \end{equation}
 where $t$ is the time, $\phi$ is the longitude, $d$ is the day of the year, d$_{off}$ an offset, and $n$ the number of days per year. The global atmospheric energy balance is improved by re-feeding the kinetic energy losses due to surface friction and horizontal and vertical momentum diffusion \citep{Lucarini2010}.  A diagnostic of the entropy budget is available \cite{Fraedrich2008}. Our average energy bias on the energy budget is smaller than 0.5~\wm in all simulations, which it is achieved locally by an instantaneous heating of the air \citep{Lucarini2011}.\\

 \begin{figure*}[ht]\centering  
	\includegraphics[width=\textwidth]{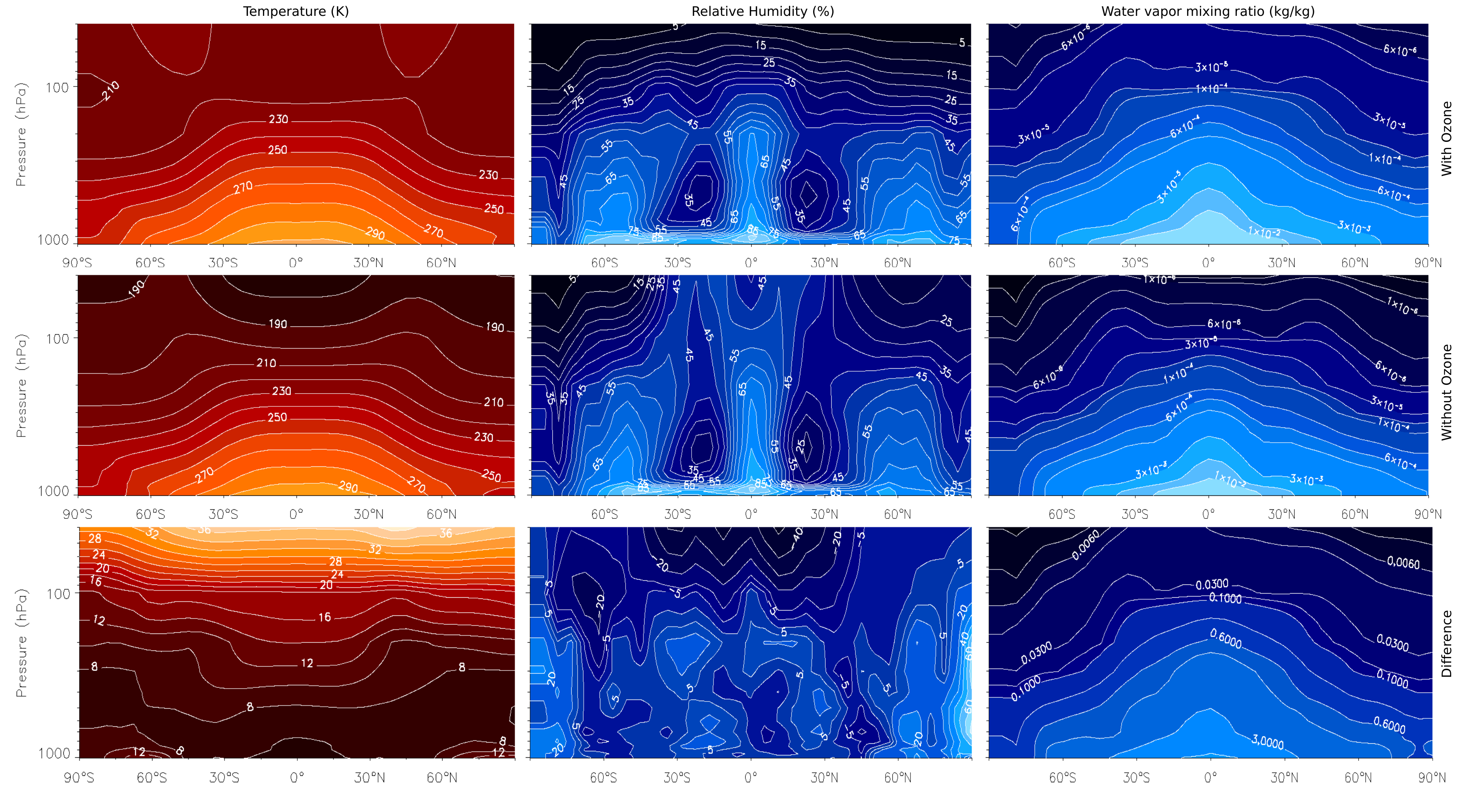}
	\caption{Zonal mean temperature (K), relative humidity (\%), and water vapor mixing ratio (kg\,kg$^{-1}$) with ozone (top) and without ozone (middle), and their difference (bottom) at the present solar irradiance ($TSI$=1361~\wmn).\label{pre}}
\end{figure*} 
\begin{figure*}[ht]\centering  
	\includegraphics[width=\textwidth]{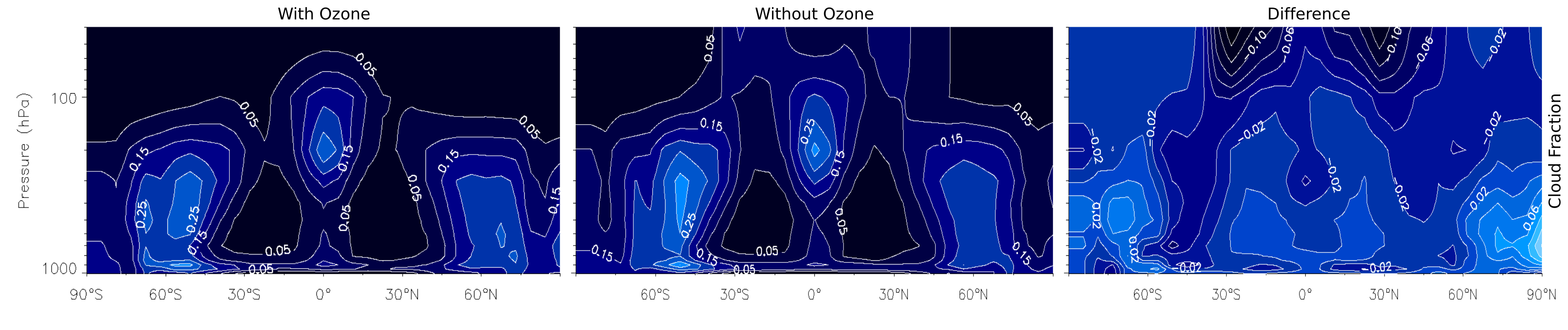}
	\caption{Zonal mean cloud fraction for the present Earth with ozone (left) and without ozone (middle), and their difference (right) at the present solar irradiance (TSI=1361~\wmn).\label{cld}}
\end{figure*}

We have used a T21 horizontal resolution ($\sim$5.6$^{\circ}$$\times$5.6$^{\circ}$ on a gaussian grid) and 18 vertical levels with the uppermost level at 40~hPa. This resolution enables to have an accurate representation of the large scale circulation features and the global thermodynamical properties of the planet \citep{Pascale2011}. The surface energy budget has been calculated as,
\begin{equation}
\Delta E= F^{net}_{SW}-F^{net}_{LW}-F^{net}_{LH}-F^{net}_{SH}-\rho_{w}L_{f}v_{SM}
\end{equation}
where $F^{net}_{SW}$ is the net shortwave radiative flux, $F^{net}_{LW}$ is the net longwave radiative flux, $F^{net}_{LH}$ is the latent heat flux, $F^{net}_{SH}$ is the sensible heat flux, $\rho_{w}$ is the density of water, $L_{f}$ is the latent heat of fusion, and $v_{SM}$ is the snow melt. The surface is in equilibrium at every state, with an energy budget $<$0.02~\wmn.
We simulate two cases: an Earth analog with the present atmospheric ozone concentration and another without ozone. We run simulations at the present solar irradiance (1361~\wmn) for both cases, and then we increase the solar forcing until the atmosphere becomes opaque. Each simulation has a length of 100 years to ensure that the system achieves the equilibrium well before the end of the run and the statistical results are averaged over the last 30 years in order to rule out the presence of transient effects.\\

\begin{figure*}
	\centering 
	\begin{minipage}{0.43\linewidth}         
		\includegraphics[width=\linewidth]{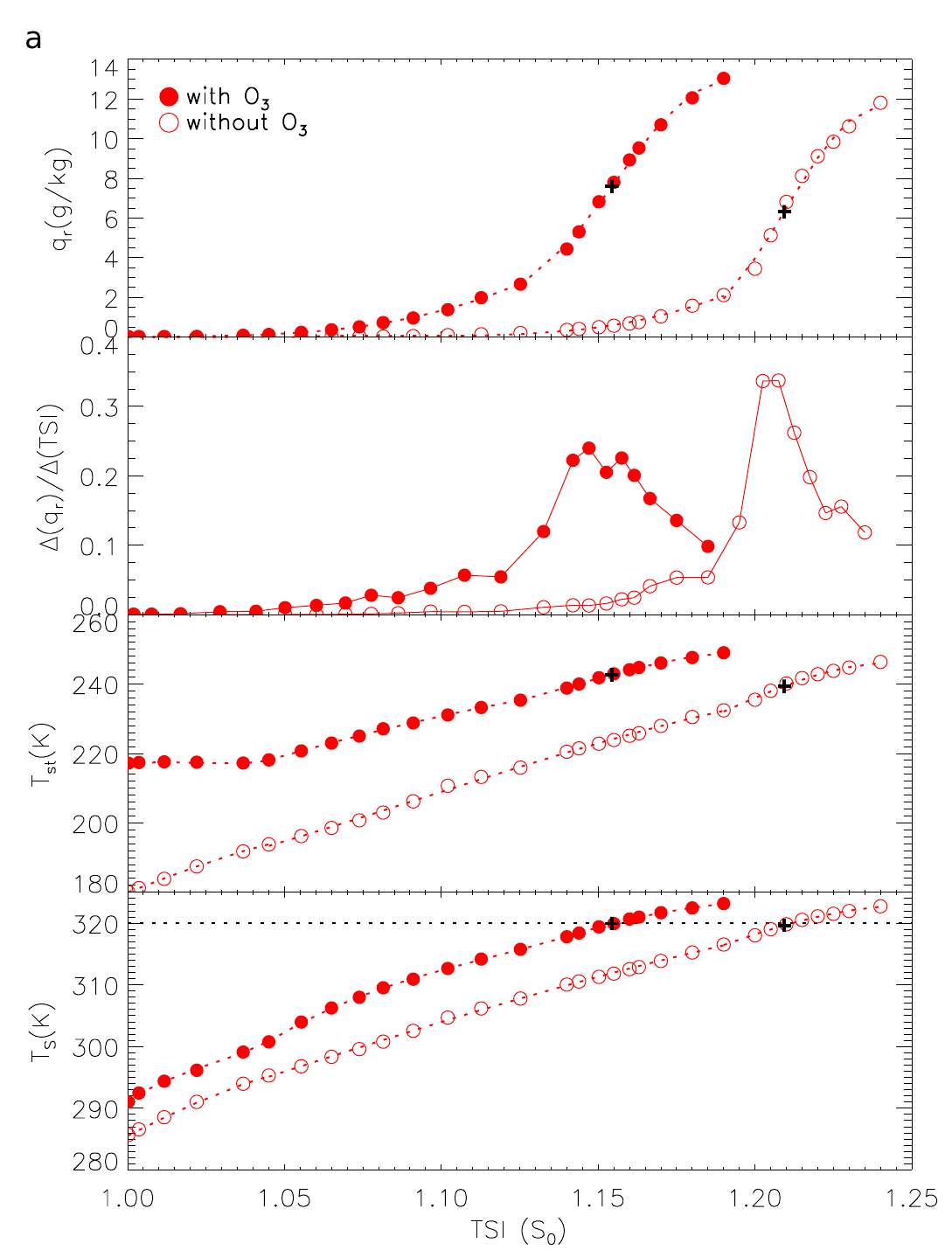}    
	\end{minipage}
	\hspace{0.05\linewidth}
	\begin{minipage}{0.43\linewidth}
		\includegraphics[width=\linewidth]{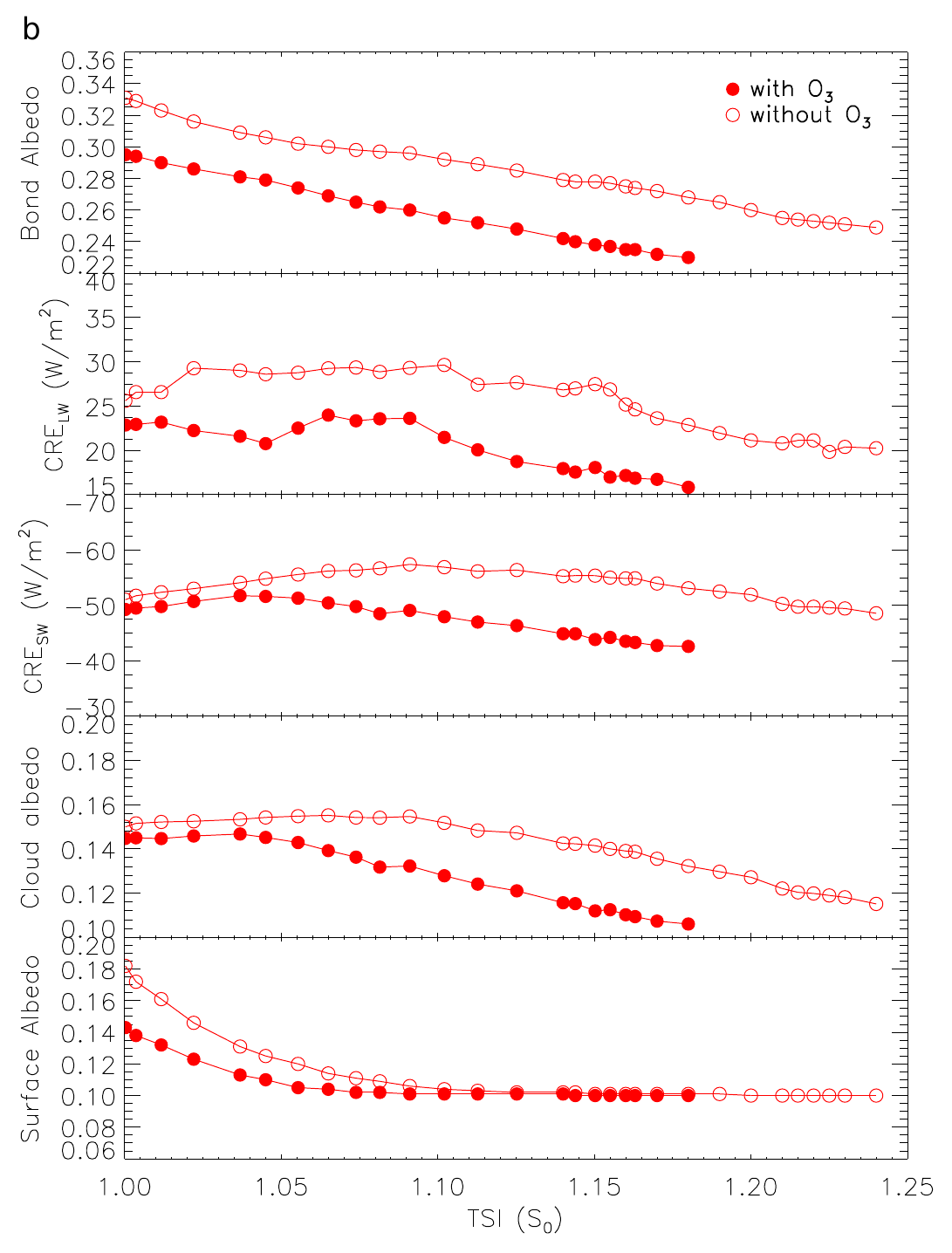}   
	\end{minipage}
	\caption{Climatic variables as a function of solar irradiance with and without $O_{3}$. a) From top to bottom: Water vapor mixing ratio ($q_{r}$) at 40~hPa, its variation with the solar irradiance, temperature (T$_{st}$) at 40~hPa, and surface temperature ($T_{S}$). The inflection points (crosses) of the polynomial approximations (dotted red lines) of the series indicate the moist greenhouse threshold in each case. b) From top to bottom: Bond albedo, cloud radiative effect ($CRE$) for the longwave (LW) radiation, $CRE$ for the shortwave (SW) radiation, cloud albedo, and surface albedo. \label{kast_cre}}  
	\vspace{2mm}
	\includegraphics[width=\textwidth]{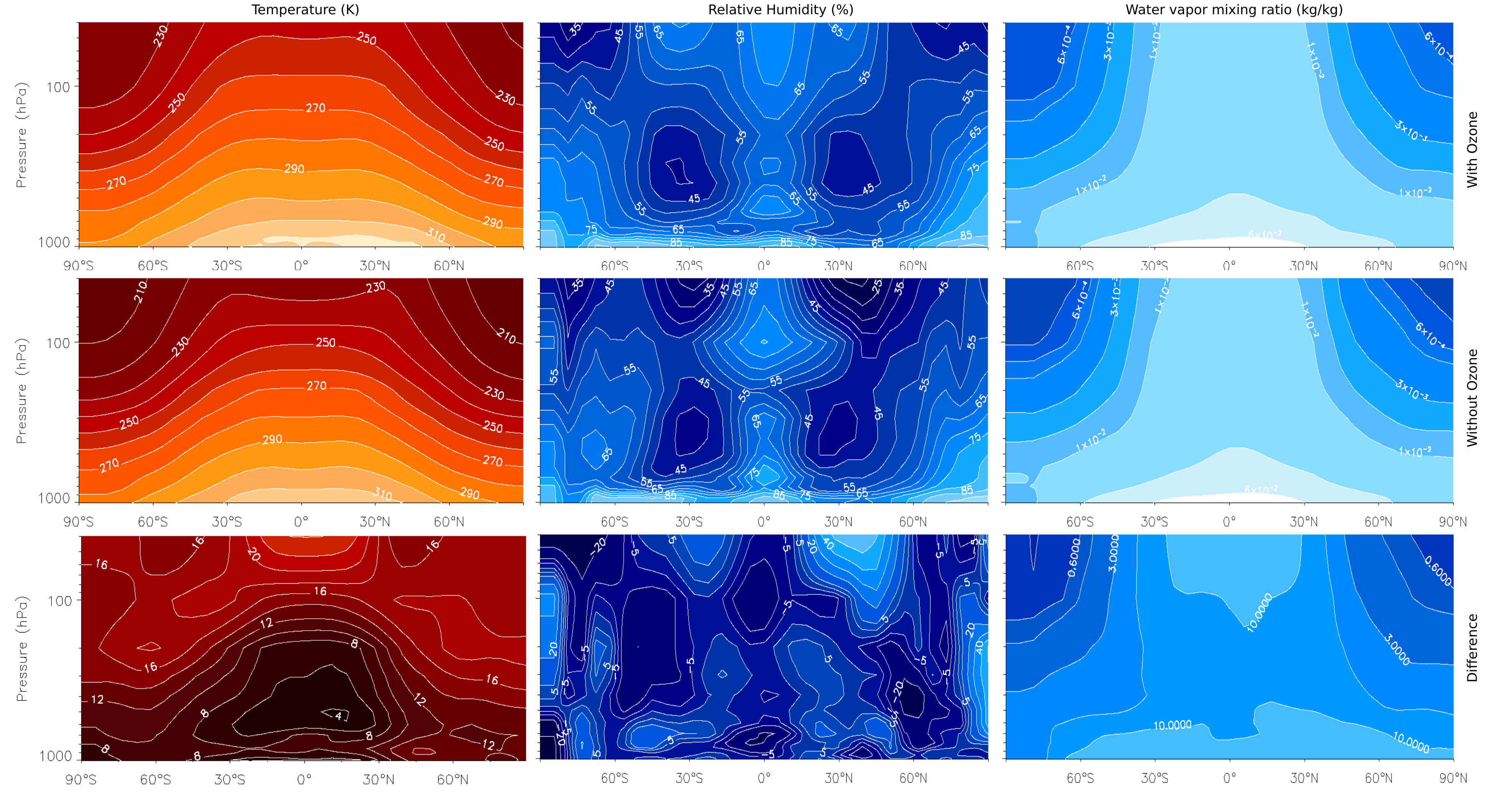}
	\caption{Zonal mean temperature (K), relative humidity (\%), and water vapor mixing ratio (kg\,kg$^{-1}$) with ozone (top) and without ozone (middle), and their difference (bottom) at a total solar irradiance of 1572~\wmn. The simulation with ozone reaches the moist greenhouse state at 1572~\wmn, while the simulation without ozone reaches this state at 1647~\wmn.\label{mgt}}
\end{figure*}

The total solar irradiance (TSI) and the concentrations of $CO_{2}$ and O$_{3}$ are inputs in the model. The surface temperature ($T_{S}$) is calculated as the global mean of the near surface air temperature. The effective temperature is calculated as the global mean of the radiative temperature at the top-of-the-atmosphere (TOA), $T_{eff}=(F^{TOA}_{LW}/\sigma)^{1/4}$, where $F^{TOA}_{LW}$ is the outgoing longwave radiation at TOA and $\sigma$ is the Stefan-Boltzmann constant. The Bond albedo is calculated as $A=1-(4\,F^{TOA}_{SW}/S_{0})$, where $F^{TOA}_{SW}$ is the reflected radiation at TOA and $S_{0}$ is the solar constant. The normalized greenhouse parameter is determined as $g_{n}=1-(T_{eff}/T_{S})^{4}$. The cloud radiative effect is calculated as the difference between the upward flux for clear-sky and for all-sky conditions $CRE=F^{up}_{clear-sky}-F^{up}_{all-sky}$ for both SW and LW ranges. The global mean temperature $T_{40}$ and water vapor mixing ratio $q_{r}$ in the stratosphere are computed at 40~hPa ($\sim$25~km). This level corresponds to an altitude about 25-30~km on the present conditions (in the mid-stratosphere), and it represents a compromise between the concentration and the dissociation of O$_{2}$, H$_{2}$O, and O$_{3}$. \citep[e.g.][]{Garcia1983, Fioletov2008}. The standard deviation is one order of magnitude lower than the values of the results presented in Table~\ref{tab1}.\\

In order to estimate the MGT, we derive the polynomial approximation of the water vapor mixing ratio of the simulation series of each case and we calculate the inflection point of the curve. The equivalent distance ($D$) in our present Solar System of the moist greenhouse limit is derived as $D= (S_{0}/TSI_{MG})^{(1/2)}$, where $D$ is expressed in astronomical units and $TSI_{MG}$, the irradiance at the MGT, is a multiple of the solar constant (\son=1361~\wmn).

\section{Results}\label{res}

First, we compare PlaSim present Earth's climate (including ozone) with the European Centre for Medium-Range Weather Forecasts (ECMWF) climate reanalysis data (ERA)\footnote{\url{http://www.ecmwf.int/en/research/climate-reanalysis/browse-reanalysis-datasets}}, which provides a consistent representation of the current climate of the Earth. PlaSim simulations are at a steady state, contrary to reanalysis data. Nonetheless, since the current climate change is relatively slow, these comparisons are meaningful, presenting the bias of the model with respect to present Earth conditions. We use ERA-20CM flux data \citep{Hersbach2015} to calculate the global surface temperature, the effective temperature, the Bond albedo of the planet, and the efficient emissivity of the atmosphere for the thermal radiation. The stratospheric temperature and the water mixing ratio have been extracted from ERA-20C data \citep{Poli2016}. The surface temperature, the effective temperature, the stratospheric temperature, and the albedo differ by less than 1\% from ERA data at the same CO$_{2}$ concentration (388 ppm) and total solar irradiance (TSI=1361~\wmn) (Table~\ref{tab1}). The tropopause lies at 200~hPa in ERA and PlaSim, the stratospheric temperatures differ by 1~K, but the water mixing ratios are 2.3$\times$10$^{-3}$~\gkg and 7.4$\times$10$^{-3}$~\gkgn, respectively. In general, PlaSim CRE, cloud albedo, surface albedo and the cloud cover of the present Earth are similar to ERA values (Figs.~\ref{prof_era} and \ref{cre_era}). In addition, we have simulated a doubling of the preindustrial CO$_{2}$ concentration (from 280~ppm to 560~ppm), in order to measure the model response to solar forcing (Table~\ref{tab1}, rows c and d). We obtain an equilibrium climate sensitivity of 2.1~K and a climate feedback parameter of 1.75~\wmn K$^{-1}$, which are within the range of values estimated by the IPCC reports \citep[e.g.][]{Bindoff2013} and other recent estimations \citep[e.g.][]{Forster2016}. 

Second, we compare the results with and without atmospheric ozone. Under the present solar irradiance, ozone warms the stratosphere in the region above 200~hPa circa. The resulting temperature inversion limits convection from penetrating above that height (Fig.~\ref{pre}). In the absence of ozone, however, the stratosphere is colder and has a lower water vapor mixing ratio at the same level (40~hPa). The pronounced inversion temperature in the stratosphere does not appear, and the decreasing temperature gradient extends to higher altitudes, allowing clouds to form higher in the atmosphere. Condensation will occur when relative humidity (RH) reaches 100\% or when the water vapor pressure equals the saturation vapor pressure. The RH increases when the air is cooled, since the specific humidity remains constant and the saturation vapor pressure decreases with decreasing temperature. Therefore, RH and cloud condensation are enhanced, while temperature and convection have lower values than in the presence of ozone (Fig.~\ref{pre}, Fig.~\ref{cld}, and Table~\ref{tab1}, rows a and b). The colder surface increases the production of ice in the poles (Fig.~\ref{pre}, the difference measured at 273~K is 20$^{\circ}$ in latitude), which in turn decreases water evaporation and makes the atmosphere drier. The Bond albedo is larger due to the increase of the surface albedo and in a less degree to the increase of the cloud albedo (Fig.~\ref{kast_cre}b), especially from equatorial convective clouds that have a larger radiative effect in both the SW and the LW ranges. In the absence of ozone, the planet reflects more incoming solar radiation, it absorbs more outgoing radiation, and the lower temperatures produce a positive ice-albedo feedback that increases the cooling of the planet.\\

%%%%%%%%
%TABLES 2 y 3
%%%%%%%%%

\begin{table*}[ht]
	\begin{minipage}{\textwidth}
		\begin{tabularx}{1.\textwidth}{| c |  c  c  l c   c   c   c  l  c   c   c  }\cline{1-10}
			\multicolumn{1}{|p{0.5cm}}{}&\multicolumn{1}{p{3.5cm}}{\centering Model} &\multicolumn{1}{|p{0.5cm}}{\centering O$_{3}$}&\multicolumn{1}{|p{0.7cm}}{\centering  Atmos. State} &\multicolumn{1}{p{2cm}}{\centering Criteria} &\multicolumn{1}{|p{1cm}}{\centering TSI (Wm$^{-2}$)} &\multicolumn{1}{p{1cm}}{\centering D (AU)} & \multicolumn{1}{|p{1cm}}{\centering A$_{S}$} & \multicolumn{1}{p{0.5cm}}{\centering Clouds} & \multicolumn{1}{p{3.5cm}|}{\centering Convective Scheme} \\\cline{1-10}
			
			\multirow{2}*{1D} & \multicolumn{1}{p{3.5cm}}{\centering \citet{Kasting1993}} &\multicolumn{1}{|p{0.5cm}}{\centering no}  &\multicolumn{1}{|p{0.7cm}}{\centering MG} &\multicolumn{1}{p{2cm}}{\centering q$_{r}$=3~\gkg} &\multicolumn{1}{|p{1cm}}{\centering 1496} &\multicolumn{1}{p{1cm}}{\centering 0.95}&\multicolumn{1}{|p{1cm}}{\centering 0.22 }&\multicolumn{1}{p{0.5cm}}{\centering no}&\multicolumn{1}{p{3.5cm}|}{\centering no}\\
			
			&\multicolumn{1}{p{3.5cm}}{\centering \citet{Kopparapu2013}}&\multicolumn{1}{|p{0.5cm}}{\centering no}  &\multicolumn{1}{|p{0.7cm}}{\centering MG} &\multicolumn{1}{p{2cm}}{\centering q$_{r}$=3~\gkg} &\multicolumn{1}{|p{1cm}}{\centering 1380} &\multicolumn{1}{p{1cm}}{\centering 0.99}&\multicolumn{1}{|p{1cm}}{\centering 0.22}&\multicolumn{1}{p{0.5cm}}{\centering no}&\multicolumn{1}{p{3.5cm}|}{\centering no}\\\cline{1-10}
			
			\multirow{3}*{3D} &\multicolumn{1}{p{3.5cm}}{\centering \citet{Leconte2013}} &\multicolumn{1}{|p{0.5cm}}{\centering no}  &\multicolumn{1}{|p{0.7cm}}{\centering RG} &\multicolumn{1}{p{2cm}}{\centering q$_{r}$=3~\gkg} &\multicolumn{1}{|p{1cm}}{\centering 1500} &\multicolumn{1}{p{1cm}}{\centering 0.95}&\multicolumn{1}{|p{1cm}}{\centering variable}&\multicolumn{1}{p{0.5cm}}{\centering yes}&\multicolumn{1}{p{3.5cm}|}{\centering convective adjustment}\\
			
			&\multirow{2}*{\citet{Wolf2015}$^{i}$} &\multicolumn{1}{|p{0.5cm}}{\centering no }&\multicolumn{1}{|p{0.7cm}}{\centering MG} &\multicolumn{1}{p{2cm}}{\centering q$_{r}$=3~\gkg} &\multicolumn{1}{|p{1cm}}{\centering 1620}  &\multicolumn{1}{p{1cm}}{\centering 0.92}&\multicolumn{1}{|p{1cm}}{\centering variable}&\multicolumn{1}{p{0.5cm}}{\centering yes}&\multicolumn{1}{p{3.5cm}|}{\centering mass-flux}\\
			
			& &\multicolumn{1}{|p{0.5cm}}{\centering no}&\multicolumn{1}{|p{0.7cm}}{\centering MG} &\multicolumn{1}{p{2cm}}{\centering c.s peak$^{ii}$} &\multicolumn{1}{|p{1cm}}{\centering 1531}  &\multicolumn{1}{p{1cm}}{\centering 0.94}&\multicolumn{1}{|p{1cm}}{\centering variable}&\multicolumn{1}{p{0.5cm}}{\centering yes}&\multicolumn{1}{p{3.5cm}|}{\centering mass-flux}\\\cline{1-10}
			
			\multirow{2}*{3D}& \multirow{2}*{This paper} & \multicolumn{1}{|p{0.5cm}}{\centering no} &\multicolumn{1}{|p{0.7cm}}{\centering  MG}&\multicolumn{1}{p{2cm}}{\centering q$_{r}$ increase$^{ii}$}&\multicolumn{1}{|p{1cm}}{\centering 1647} &\multicolumn{1}{p{1cm}}{\centering 0.91}&\multicolumn{1}{|p{1cm}}{\centering variable}&\multicolumn{1}{p{0.5cm}}{\centering yes}&\multicolumn{1}{p{3.5cm}|}{\centering moist convergence}\\
			
			& &\multicolumn{1}{|p{0.5cm}}{\centering yes}	&\multicolumn{1}{|p{0.7cm}}{\centering  MG}&\multicolumn{1}{p{2cm}}{\centering q$_{r}$ increase$^{ii}$}  &\multicolumn{1}{|p{1cm}}{\centering 1572} &\multicolumn{1}{p{1cm}}{\centering 0.93}&\multicolumn{1}{|p{1cm}}{\centering variable}&\multicolumn{1}{p{0.5cm}}{\centering yes}&\multicolumn{1}{p{3.5cm}|}{\centering moist convergence}\\\cline{1-10}
		\end{tabularx}
		%\tablecomments{\footnotesize {$^{i}$\cite{Wolf2015} identifies two possible moist greenhouse states: one at 1620~\wmn, corresponding to the water vapor mixing ratio (q$_{r}$) of 3~\gkg\, given by \cite{Kasting1993}, and other at 1532~\wmn, corresponding to the point of maximum climate sensitivity (c.s.). $^{ii}$Our q$_{r}$ increase coincides with the climate sensitivity peak, the second criteria used by \cite{Wolf2015}.}}
		\caption{Comparison between models: ozone (O$_{3}$), atmospheric state (moist greenhouse or runaway greenhouse), criteria used to define the atmospheric state, total solar irradiance (TSI), distance, surface albedo (A$_{S}$), cloud scheme, and convective scheme. N\scriptsize{OTE}--- \footnotesize {$^{i}$\cite{Wolf2015} identifies two possible moist greenhouse states: one at 1620~\wmn, corresponding to the water vapor mixing ratio (q$_{r}$) of 3~\gkg\, given by \cite{Kasting1993}, and other at 1532~\wmn, corresponding to the point of maximum climate sensitivity (c.s.). $^{ii}$Our q$_{r}$ increase coincides with the climate sensitivity peak, the second criteria used by \cite{Wolf2015}.}}\label{tab2}
		%\end{table*}
		\smallbreak
		%\begin{table*}[hbt!]
		\begin{tabularx}{1.\textwidth}{c  c l c   c   c  l c  c   c   c  l c   c   }
			\multicolumn{11}{l}{\hspace{1.5cm}$\emph{The present Earth's climate}$}\\\cline{2-11}
			&\multicolumn{1}{|p{1.5cm}}{\centering Model} &\multicolumn{1}{|p{1cm}}{\centering $O_{3}$} & \multicolumn{1}{p{1.8cm}}{\centering $TSI$(\wmn)} &\multicolumn{1}{p{2cm}}{\centering [$CO_{2}$](ppm)} &\multicolumn{1}{|p{1cm}}{\centering $T_{S}$(K)}  &\multicolumn{1}{p{1.cm}}{\centering $T_{eff}$(K)} & \multicolumn{1}{p{1cm}}{\centering A} & \multicolumn{1}{p{0.5cm}}{\centering $g_{n}$}& \multicolumn{1}{|p{1cm}}{\centering T$_{40}$(K)} &\multicolumn{1}{p{2cm}|}{\centering $q_{r}$(\gkgn)} \\\cline{1-11}
			
			\multicolumn{1}{|c}{\centering }&\multicolumn{1}{|c}{ERA}  	      & \multicolumn{1}{|c}{yes}& \multicolumn{1}{c}{1361} & 388  & \multicolumn{1}{|c}{289.1}  &\multicolumn{1}{c}{255.3}  & 0.294 & 0.392  & \multicolumn{1}{|c}{216} & \multicolumn{1}{c|}{2.3$\times$10$^{-3}$}\\\cline{1-11}
			\multicolumn{1}{|c}{\centering a}&\multicolumn{1}{|c}{LMDZ$^{i}$}   & \multicolumn{1}{|c}{no} & \multicolumn{1}{c}{1365} & 376  & \multicolumn{1}{|c}{282.8}  &\multicolumn{1}{c}{253.8} & 0.311  & 0.351  & \multicolumn{1}{|c}{170} & \multicolumn{1}{c|}{1$\times$10$^{-5}$}\\
			\multicolumn{1}{|c}{\centering b}&\multicolumn{1}{|c}{PlaSim} & \multicolumn{1}{|c}{no}& \multicolumn{1}{c}{1365}  & 376  & \multicolumn{1}{|c}{285.0}  & \multicolumn{1}{c}{252.3} & 0.330  & 0.397 & \multicolumn{1}{|c}{181} & \multicolumn{1}{c|}{6.3$\times$10$^{-4}$} \\
			\multicolumn{1}{|c}{\centering c}&\multicolumn{1}{|c}{PlaSim} 		& \multicolumn{1}{|c}{yes} & \multicolumn{1}{c}{1365}  & 376 & \multicolumn{1}{|c}{291.6} & \multicolumn{1}{c}{255.6} & 0.295  & 0.420 & \multicolumn{1}{|c}{218} & \multicolumn{1}{c|}{9.0$\times$10$^{-3}$} \\\cline{1-11}
			\multicolumn{1}{|c}{\centering d}&\multicolumn{1}{|c}{CAM4$^{ii}$}  & \multicolumn{1}{|c}{no}& \multicolumn{1}{c}{1361}  & 367  & \multicolumn{1}{|c}{289.1}  &\multicolumn{1}{c}{252.0} & 0.329 & 0.423  & \multicolumn{1}{|c}{170} & \multicolumn{1}{c|}{1$\times$10$^{-5}$}\\
			\multicolumn{1}{|c}{\centering e}&\multicolumn{1}{|c}{PlaSim} & \multicolumn{1}{|c}{no}& \multicolumn{1}{c}{1361}& 367 & \multicolumn{1}{|c}{283.6}  & \multicolumn{1}{c}{251.7} & 0.335  & 0.392 & \multicolumn{1}{|c}{180} & \multicolumn{1}{c|}{4.9$\times$10$^{-4}$}\\
			\multicolumn{1}{|c}{\centering f}&\multicolumn{1}{|c}{PlaSim} 		& \multicolumn{1}{|c}{yes}& \multicolumn{1}{c}{1361} & 367  & \multicolumn{1}{|c}{290.8} & \multicolumn{1}{c}{255.2} & 0.297  & 0.417 & \multicolumn{1}{|c}{217} & \multicolumn{1}{c|}{7.3$\times$10$^{-3}$} \\\cline{1-11}
		\end{tabularx}
		%\tablecomments{\footnotesize {$^{i}$ \citet{Leconte2013}; $^{ii}$ \citet{Wolf2015}}}
		\caption{Comparison of Earth's present state in ERA, \citet{Leconte2013}, \citet{Wolf2015}, and PlaSim. The ozone ($O_{3}$) concentration, the total solar irradiance ($TSI$), and the $CO_{2}$ concentration (in ppm) are initial conditions. The surface temperature ($T_{S}$), the effective temperature ($T_{eff}$), the Bond albedo (A), the normalized greenhouse parameter ($g_{n}$) calculations are explained in Section~\ref{mod}. The temperature (T$_{40}$) and the water vapor mixing ratio ($q_{r}$) are both measured at a pressure level of 40~hPa. N\scriptsize{OTE}--- \footnotesize {$^{i}$ \citet{Leconte2013}; $^{ii}$ \citet{Wolf2015}}}\label{tab3}
	\end{minipage}
\end{table*} 
%%%%%%

The response to the increasing solar irradiance is also amplified by the ice-albedo feedback: as the surface temperature rises, the ice and snow melt, decreasing the surface albedo and the Bond albedo (Fig.~\ref{kast_cre}a). As a result, the planet absorbs more solar radiation (Fig.~\ref{kast_cre}b). The water vapor increases, absorbing more infrared radiation and warming the atmosphere, which in turn raises the surface temperature. The surface albedo reaches its minimum when the globally averaged surface temperatures is higher than 300~K, due to the complete melt of the ice and snow of the planet. This is attained at a TSI$\sim$1.05~\so (1429~\wmn) with ozone and at a TSI$\sim$1.10~\so (1497~\wmn) without ozone, since in the absence of its warming effect, the planet needs more solar radiation to obtain the same surface temperature. These energies correspond to a distance of 0.975~au and 0.953~au in the present Solar System. The global melting of ice and snow has important effects on the large-scale climate. As discussed in \citet{Boschi2013}, the hydrological cycle becomes the predominant climatic factor: the moist convection increases and the meridional heat transport is only controlled by the amount of water vapor in the atmosphere. Our results show that, as a result of the enhanced moist convection, water vapor starts to increase in the stratosphere once the ice and snow have melted (Fig.~\ref{kast_cre}a).\\

%%%%%%
%TAB4
%%%%%%%%%%%

\begin{table*}[ht]
	\centering
	\begin{center}
	\begin{tabularx}{1.\textwidth}{|c c  l  c   l  c   c   c   c  l  c  c }\cline{1-9}
		\multicolumn{9}{|c|}{\hspace{1.5cm}\emph{$TSI=1361$\wmn}}\\\cline{1-9}
		\multicolumn{1}{|p{1cm}}{\centering Model} &\multicolumn{1}{|p{3cm}}{\centering Convection scheme} &\multicolumn{1}{|p{1.5cm}}{\centering $O_{3}$} &\multicolumn{1}{|p{1.5cm}}{\centering $T_{S}$(K)}  &\multicolumn{1}{p{1.5cm}}{\centering $T_{eff}$(K)} & \multicolumn{1}{p{1.5cm}}{\centering A} & \multicolumn{1}{p{1.5cm}}{\centering $g_{n}$} & \multicolumn{1}{|p{1.5cm}}{\centering T$_{st}$(K)} &\multicolumn{1}{p{1.5cm}|}{\centering $q_{r}$(\gkgn)} \\\cline{1-9}		
		\multicolumn{1}{|c}{ERA} & \multicolumn{1}{|c}{-} & \multicolumn{1}{|c}{yes} &  \multicolumn{1}{|c}{289.1}&\multicolumn{1}{c}{255.3} & 0.294 & 0.392 & \multicolumn{1}{|c}{216} &\multicolumn{1}{c|}{2.3$\times$10$^{-3}$}\\\cline{1-9}\cline{1-9}
		
		\multirow{2}*{PlaSim}& \multicolumn{1}{|c}{Kuo} & \multicolumn{1}{|c}{yes} &  \multicolumn{1}{|c}{291.0}&\multicolumn{1}{c}{255.2} & 0.296 & 0.419 & \multicolumn{1}{|c}{217} &\multicolumn{1}{c|}{7.4$\times$10$^{-3}$}\\
		&\multicolumn{1}{|c}{Betts-Miller}  & \multicolumn{1}{|c}{yes} & \multicolumn{1}{|c}{292.2}&\multicolumn{1}{c}{254.9} & 0.298 & 0.424 & \multicolumn{1}{|c}{217} &\multicolumn{1}{c|}{7.1$\times$10$^{-3}$}\\\cline{1-9}
		
		\multirow{2}*{PlaSim}&\multicolumn{1}{|c}{Kuo} & \multicolumn{1}{|c}{no} & \multicolumn{1}{|c}{284.0} &\multicolumn{1}{c}{251.8} & 0.334 & 0.394 & \multicolumn{1}{|c}{180} &\multicolumn{1}{c|}{5.0$\times$10$^{-4}$} \\
		\multicolumn{1}{|c}{}&\multicolumn{1}{|c}{Betts-Miller}  & \multicolumn{1}{|c}{no} & \multicolumn{1}{|c}{286.5}&\multicolumn{1}{c}{252.9} & 0.280 & 0.393 & \multicolumn{1}{|c}{185} &\multicolumn{1}{c|}{1.0$\times$10$^{-5}$} \\\cline{1-9}
		\multicolumn{1}{|c}{CAM4} & \multicolumn{1}{|c}{mass-flux}  & \multicolumn{1}{|c}{no} & \multicolumn{1}{|c}{289.1}&\multicolumn{1}{c}{252.0} & 0.329 & 0.423 & \multicolumn{1}{|c}{170} &\multicolumn{1}{c|}{1.0$\times$10$^{-5}$}\\\cline{1-9}		
		\multicolumn{9}{|c|}{\hspace{1.5cm}$\emph{$TSI=1572~Wm^{-2}$}$}\\\cline{1-9}
		\multicolumn{1}{|p{1cm}}{Model}& \multicolumn{1}{|p{3cm}}{\centering Convection scheme} &\multicolumn{1}{|p{1.5cm}}{\centering $O_{3}$} &\multicolumn{1}{|p{1.5cm}}{\centering $T_{S}$(K)}  &\multicolumn{1}{p{1.5cm}}{\centering $T_{eff}$(K)} & \multicolumn{1}{p{1.5cm}}{\centering A} & \multicolumn{1}{p{1.5cm}}{\centering $g_{n}$} & \multicolumn{1}{|p{1.5cm}}{\centering T$_{st}$(K)} &\multicolumn{1}{p{1.5cm}|}{\centering $q_{r}$(\gkgn)}\\\cline{1-9}
		
		\multirow{2}*{PlaSim}&\multicolumn{1}{|c}{Kuo} & \multicolumn{1}{|c}{yes} &\multicolumn{1}{|c}{319.9}&\multicolumn{1}{c}{269.5} & 0.238 & 0.496 &\multicolumn{1}{|c}{243} &\multicolumn{1}{c|}{7.8} \\
		\multicolumn{1}{|c}{}&\multicolumn{1}{|c}{Betts-Miller}  & \multicolumn{1}{|c}{yes} &  \multicolumn{1}{|c}{321.6} &\multicolumn{1}{c}{270.5} & 0.237 & 0.499 & \multicolumn{1}{|c}{223} &\multicolumn{1}{c|}{0.6} \\\cline{1-9}
		
		\multirow{2}*{PlaSim}&\multicolumn{1}{|c}{Kuo} & \multicolumn{1}{|c}{no} &  \multicolumn{1}{|c}{311.8}  &\multicolumn{1}{c}{266.1} & 0.277 & 0.470 &\multicolumn{1}{|c}{224} &\multicolumn{1}{c|}{0.6} \\
		\multicolumn{1}{|c}{}&\multicolumn{1}{|c}{Betts-Miller}  & \multicolumn{1}{|c}{no} & \multicolumn{1}{|c}{316.3}&\multicolumn{1}{c}{266.9} & 0.274 & 0.493 & \multicolumn{1}{|c}{214} &\multicolumn{1}{c|}{0.2} \\\cline{1-9}		
		\multicolumn{1}{|c}{CAM4} & \multicolumn{1}{|c}{mass-flux}  & \multicolumn{1}{|c}{no} & \multicolumn{1}{|c}{337.1}&\multicolumn{1}{c}{226.4} & 0.270 & 0.610 & \multicolumn{1}{|c}{240} &\multicolumn{1}{c|}{[3, 10]}\\\cline{1-9}\cline{1-9}
		
 \end{tabularx}
	
	\caption{Comparison of PlaSim simulations with Kuo and Betts-Miller convection schemes, with ERA reanalysis data, and \cite{Wolf2015} results using CAM4 with a mass-flux scheme. From top to bottom: Mean global data at 1361~\wm (Earth's present state) and at 1572~\wm (the moist greenhouse threshold in the case with atmospheric ozone): ozone ($O_{3}$), surface temperature ($T_{S}$), effective temperature ($T_{eff}$), Bond albedo (A), normalized greenhouse parameter ($g_{n}$), stratospheric temperature (T$_{st}$), and water vapor mixing ratio ($q_{r}$) at a pressure level 40~hPa.} \label{tab4}
	\end{center}
\end{table*} 

%%%%%%%%%%%

We obtain a peak of the water vapor mixing ratio between 1.14~\so and 1.16~\son, in the case with ozone, and a peak between 1.20~\so and 1.21~\son, in the case without ozone. In order to calculate the MGT, we calculate the polynomial approximation of the water vapor mixing ratio series and its inflection point. Our results (Figs.~\ref{prof_era}, \ref{mgt}, and Table~\ref{tab2}) indicate that the Earth would reach the MGT under a solar irradiance of about 1572~\wm (1.155~\son).  At this irradiance, the water mixing ratio is about 10 times larger than without ozone, the temperature is 8 degrees warmer at the surface, and about 20 degrees warmer at 40~hPa. Without ozone, the MGT occurs at a TSI$\sim$1647~\wm (1.209~\son). The solar model proposed by \citet{Bahcall2001} predicts these irradiance values in 1.6 and 2.1 billion years, respectively.  These radiation values correspond to an equivalent distance in the present Solar System of about 0.93~au with ozone and 0.91~au without ozone, indicating a shift in the inner limit of the conservative Habitable Zone due to the O$_{3}$ concentration in the atmosphere. An Earth-like planet with the present ozone concentration starts to lose its water 500 million years earlier than a similar planet without ozone in its atmosphere.

\subsection{Comparison with previous studies}\label{com}
We compare our results with earlier GCM studies on the greenhouse effect under an increasing solar forcing without ozone. \citet{Leconte2013} (here after L13) simulates an Earth-like planet with an atmosphere composed by 1 bar of N$_{2}$, a variable amount of water vapor, CO$_{2}$ concentration of 376~ppm, and an initial TSI=1365\wmn, using a modified version of the LMD Generic GCM (LMDG); and \citet{Wolf2015} (here after W15)  uses a modified version of the Community Atmosphere Model (CAM4) with a similar composition of the atmosphere, 367~ppm of CO$_{2}$, and an initial TSI=1361.27~\wm (Table~\ref{tab3}, rows a and d, respectively). Both models use a photochemical atmospheric module. L13 accounts for the fact that water vapor is a non-trace gas in a hot humid atmosphere. They do not include ozone. Therefore, their tropopause at the present state is placed above 40~hPa, and the water mixing ratio ($\sim$10$^{-5}$~\gkgn) and the temperature (170~K) at 40~hPa in both models are much lower than in ERA and PlaSim. The surface temperature is 6~K colder in L13 data than in ERA and the Bond albedo is 6\% higher. In W15, the surface temperature is similar to ERA data and an albedo 12\% higher than our planet. The surface temperature in PlaSim simulations without ozone at the present solar irradiance (Tables~\ref{tab1} and ~\ref{tab3}) are about 5~K colder than ERA. Our results are consistent with the fact that ozone warms the atmosphere, therefore in the absence of ozone, surface temperatures are colder than present Earth, in agreement with \citet{Bordi2012}.\\

Our simulations without ozone at the present state show a Bond albedo 1\% larger than in W15 and 7\% larger than in L13. The humidity in the stratosphere is about 5 times larger than in W15 and 500 times larger than in L13. At higher solar irradiances, these differences have an impact on the climate of the planet. At a TSI=1500~\wmn, the surface temperature is 312~K in W15, 310~K in PlaSim, and 335~K in L13 (L13 does not have data with a higher solar forcing). At a TSI=1572~\wmn, however, the surface temperature is 337~K in W15 and 320~K in PlaSim (Table~\ref{tab4}).

\citet{Wolf2015} identifies the MGT by a climate sensitivity peak at a TSI$\sim$1531~\wm (1.125~\son), which corresponds to a time of 1.3 billion years and a distance of 0.94~au at the present in the Solar System. We obtain a higher radiation value, which can be explained by the 5~K difference in the surface temperature at the present irradiance. Our results show that an Earth without ozone will remain habitable 770 million years longer than predicted by \citet{Wolf2015}.\\

\begin{figure*}[ht]\centering  
	\includegraphics[width=0.8\textwidth]{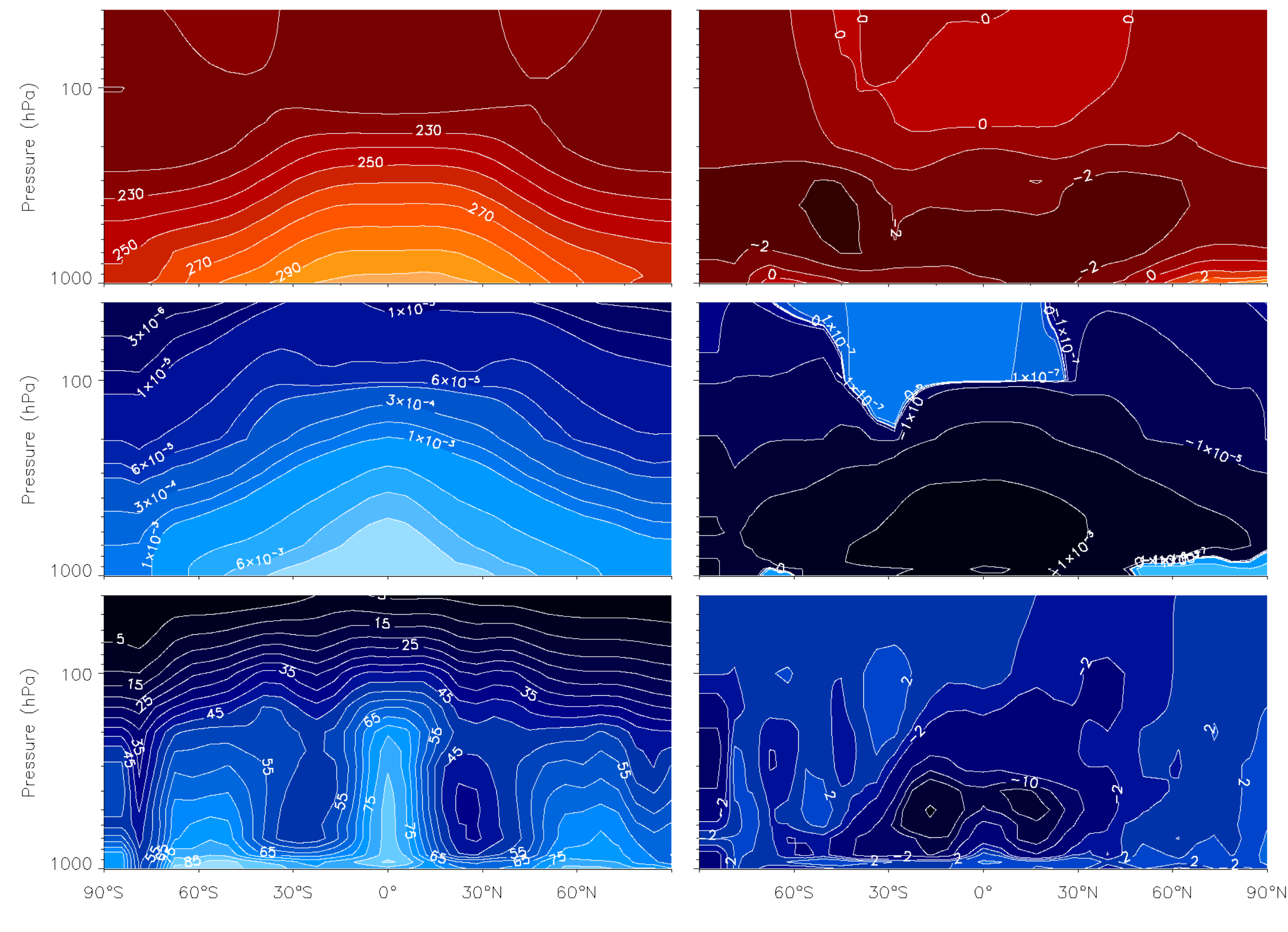}
	\caption{From top to bottom: Zonal mean temperature (K), water vapor mixing ratio (kg\,kg$^{-1}$), and relative humidity (\%) zonal maps for the present Earth simulated using the Betts-Miller moist adjustment scheme version in \cite{Frierson2007} (left) and difference using the Kuo-type scheme (Kuo minus BM).}\label{kuo_bm}
\end{figure*}

\subsection{Comparison of cumulus convection parameterization schemes}

The discrepancies between the GCM results shown above could be due to the use of different moist convection schemes (see Section~\ref{mod}). The L13 uses a convective adjustment scheme \citep{Forget1998}, PlaSim results in this article use a moist convergence scheme \citep{Kuo1965, Kuo1974}, and W15 a mass-flux scheme \citep{Zhang1995}. In general, the last two methods represent the penetrative cumulus convection and its interaction with the environment, which are important to account for the distribution of humidity in the atmosphere, while the convective adjustment scheme does not include these effects.\\

PlaSim can also use a simplified version of Betts-Miller (BM) convective adjustment scheme proposed by \citet{Frierson2007}. Here we compare PlaSim simulations under the same conditions using the BM scheme and the Kuo scheme (Table~\ref{tab4} and Fig.~\ref{kuo_bm}). The results of the simulations including ozone at the present solar irradiance are similar: the surface temperature shows a deviation of 1K and 2K, respectively in comparison to ERA data, the temperature at 40~hPa is 1K higher in both cases, and the mixing ratio is about two times the mean value shown by ERA. The results without atmospheric ozone of both schemes show a colder stratosphere than ERA and also a colder surface. There is less water vapor in the stratosphere, and as previously explained, the Bond albedo increases due to higher cloud and surface albedos. The BM scheme produces less water vapor in the stratosphere and higher temperatures than Kuo, about 2~K on the surface and 5~K in the stratosphere. At 1572~\wmn, the results including ozone show similar surface temperature, Bond albedo, effective temperature, and greenhouse parameter. However, the results in the stratosphere differ: the stratosphere is 20~K colder and about 10 times less humid with the BM scheme than with the Kuo scheme. In the case without ozone, the surface temperature is 5~K warmer, whereas the stratosphere is 10~K colder and 3 times less humid than using the Kuo scheme.

Comparing these results with the simulations in \cite{Wolf2015} using a mass-flux scheme, we note that the surface temperature at the present irradiance is equal to ERA data and 3~K higher in W15 with respect to BM. At TSI=1572~\wmn, the mean surface temperature is about 20~K higher than using the BM scheme and 25~K higher than with the Kuo scheme. We measure the temperature and the water mixing ratio in the stratosphere at the same pressure level (40~hPa) in all the cases. We want to note that the stratosphere is 30~K warmer and the water vapor is at least 10 times higher in W15 than using the BM scheme, but the results are similar to Kuo stratosphere with ozone at the same irradiance. Despite this resemblance, their climate differ the present solar irradiance, making difficult the comparison.

We conclude that at higher irradiance, the adjustment scheme give the coldest and driest stratosphere, while the moist convergence scheme shows the coldest surface temperatures. PlaSim obtains a better representation of present Earth's climate than previous studies by including ozone. In addition, a proper calibration of the model and its response to solar forcing, as well as the use of a penetrative cumulus convection scheme are essential to simulate the moist greenhouse effect. However, further research has to be made in order to clarify the divergence in temperature and humidity between cumulus convection schemes.

\section{Conclusions}\label{conc}

The moist greenhouse threshold (MGT) has been identified by a water vapor mixing ratio of about 3~\gkg in the stratosphere and a saturated troposphere by 1D models \citep{Kasting1988, Kasting1993, Kopparapu2013}. However, in 3D models, some sub-grid processes that affect humidity, such as moist convection and cloud condensation are still difficult to simulate. As a consequence, the amount of humidity in the atmosphere depends heavily on the model used, and the results for the MGT differ greatly between models \citep{Leconte2013, Wolf2015}. In addition, these 3D studies determine the MGT by using the water mixing ratio value derived by 1D models, and they do not include ozone. Ozone warms the stratosphere, modifying as well the temperature structure of the troposphere. Humidity increases due to the higher temperatures and the resulting stratification limits cloud condensation and defines the level of the tropopause.

We show that a proper calibration of the initial state, the measure of the response to solar forcing, as well as the use of a complex moist convection scheme are a key point to gain confidence in the model capabilities to adequately represent hot climates. We derive the MGT by the increase of the water mixing ratio in the stratosphere, for the first time with a 3D model. 

We find that the warming effect of ozone considerably increases the humidity of the lower atmosphere and the surface temperature. Due to the higher temperature, both the surface albedo and the planetary albedo are lower than in the case without ozone, and the greenhouse effect is enhanced. As a consequence, the MGT (TSI=1.155~\son=1572~\wmn, which corresponds to a distance of 0.93~au in the Solar System) is reached at a lower solar irradiance than in simulations without ozone (TSI=1.209~\son=1647~\wmn, which corresponds to 0.91~au), showing that, although ozone is not abundant in the atmosphere, it has relevant effects on our planet's climate and it substantially reduces the maximum solar irradiance at which an Earth-like planet would remain habitable.\\

Ozone may have important consequences on the habitability and life on other planets. A raise in the ozone concentration due to abiotic or biotic means (e.g. the Great Oxidation event) might abruptly increase the temperature and humidity of the planet and trigger the moist greenhouse state earlier than expected. Planets may be habitable at different distances from its host star depending on their ozone concentration. This study does not include a photochemical model or a detailed stratospheric scheme to simulate the Brewer-Dobson circulation. A more complete understanding of the moist greenhouse effect requires to improve our ozone models to include the large range of effects and feedbacks between water vapor, ozone, temperature, circulation, the gradual changes in the ocean and the surface, or the evolving stellar UV irradiation. Further work is needed to explore the overall evolution of ozone and its effect on the habitability of terrestrial planets.

\acknowledgements
\makeatletter{}We thank Edilbert Kirk and Tom Shannon for their assistance with PlaSim performance. We thank Eric Wolf and J\'{e}r\'{e}my Leconte for making available CAM4 and LMDG data to us, and for the valuable discussion. The authors acknowledge support by the Simons Foundation (SCOL $\#$290357, Kaltenegger), the Carl Sagan Institute, and the Centre for the Mathematics of Planet Earth of the University of Reading.

%%%%%%%%%%%%

\hspace{0.25in}
\bibliographystyle{aasjournal}

\begin{thebibliography}{}
\expandafter\ifx\csname natexlab\endcsname\relax\def\natexlab#1{#1}\fi
\providecommand{\url}[1]{\href{#1}{#1}}
\providecommand{\dodoi}[1]{doi:~\href{http://doi.org/#1}{\nolinkurl{#1}}}
\providecommand{\doeprint}[1]{\href{http://ascl.net/#1}{\nolinkurl{http://ascl.net/#1}}}
\providecommand{\doarXiv}[1]{\href{https://arxiv.org/abs/#1}{\nolinkurl{https://arxiv.org/abs/#1}}}

%%%%%%%%%%%%%%%%%%%

	\bibitem[Abe et al.(2011)]{Abe2011} Abe, Y., Abe-Ouchi, A., Sleep, N.~H., Zahnle, K.~J. 2011. \asbio, 11, 443. \dodoi{10.1089/ast.2010.0545}

\bibitem[Arakawa and Schubert(1974)]{Arakawa1974} Arakawa, A., Schubert, W.~H. 1974.  \jas, 31, 674. \dodoi{10.1175/1520-0442(2004)017<2493:RATCPP>2.0.CO;2}

\bibitem[Arakawa(2004)]{Arakawa2004} Arakawa, A. 2004. \jcli, 17, 2493. \dodoi{10.1175/1520-0469(1974)031<0674:IOACCE>2.0.CO;2}

\bibitem[Bahcall et al.(2001)]{Bahcall2001} Bahcall, J.~N., Pinsonneault, M.~H., Basu, S. 2001. \apj, 555, 990. \dodoi{10.1086/321493}

\bibitem[Berger(1978a)]{Berger1978a} Berger, A.~L. 1978a. \jas, 35, 2362. \dodoi{0.1175/1520-0469(1978)035<2362:LTVODI>2.0.CO;2}

\bibitem[Berger(1978b)]{Berger1978b} Berger, A.~L. 1978b. QuRes, 9, 2, 139. \dodoi{10.1016/0033-5894(78)90064-9}

\bibitem[Betts(1986)]{Betts1986} Betts, A.~K. 1986. QJRMS, 112, 677. \dodoi{10.1002/qj.49711247307}

\bibitem[Bindoff et al.(2013)]{Bindoff2013} Bindoff, N.~L., Stott, P.~A., Achuta Rao, et al. 2013. In Climate Change 2013. Stocker, T.,~F., et al. eds. Cambridge University Press, 867. \dodoi{10.1017/CBO9781107415324.022}

\bibitem[Boer et al.(1984)]{Boer1984} Boer, G. J., McFarlane, N.~A., Laprise, R., et al. 1984. \ato, 22, 397. \dodoi{10.1080/07055900.1984.9649208}

\bibitem[Bordi et al.(2012)]{Bordi2012} Bordi, I., Fraedrich, K., Sutera, A., \& Zhu, X. 2012. ThApC, 109, 253. \dodoi{10.1007/s00704-011-0579-5}

\bibitem[Boschi et al.(2013)]{Boschi2013} Boschi, R., Lucarini, V., \& Pascale, S.\ 2013. \icarus, 226, 1724. \dodoi{10.1016/j.icarus.2013.03.017}

\bibitem[Canuto et al.(1983)]{Canuto1983} Canuto, V.~M., Levine, J.~S., Augustsson, T.~R., \& Imhoff, C.~L. 1983. PreR, 20, 109. \dodoi{10.1016/S0166-2635(08)70238-X}

\bibitem[Domagal-Goldman and Meadows(2010)]{Domagal2010} Domagal-Goldman, S.~D., Meadows, V.~S. 2010. Pathways Towards Habitable Planets, 430, 152.

\bibitem[Domagal-Goldman et al.(2014)]{Domagal2014} Domagal-Goldman, S.~D., Segura, A., Claire, M.~W., Robinson, T.~D., \& Meadows, V.~S. 2014. \apj, 792, 90. \dodoi{10.1088/0004-637X/792/2/90}

\bibitem[Eliasen et al.(1970)]{Eliasen1970} Eliasen, E., Machenhauer, B., \& Rasmussen, E. 1970. Report No. 2. Institute for theoretical Meteorology, Copenhaguen University, Denmark.


\bibitem[Evans et al.(1998)]{Evans1998} Evans, S.~J., Toumi, R., Harries, J.~E., et al. 1998. \jgr, 103, 8715. \dodoi{10.1029/98JD00265}


\bibitem[Fast(2006)]{Fast2006} Fast, K.~E. 2006. Ozone Abundance on Mars from infrared heterodyne spectra: I. Acquisition, retrieval, and anticorrelation with water vapor.\ Ph.D.~Thesis 3177. \dodoi{10.1016/j.icarus.2005.12.001}


\bibitem[Finney et al.(2016)]{Finney2016} Finney, D.~L., Doherty, R.~M., Wild, O., Young, P.~J., \& Butler, A. 2016. \grl, 43, 5492. \dodoi{10.1002/2016GL068825}

\bibitem[Fioletov(2008)]{Fioletov2008} Fioletov; V.~E. 2008. \ato, 46, 39. \dodoi{10.3137/ao.460103}

\bibitem[Forget et al.(1998)]{Forget1998} Forget, F., Hourdin, F., \& Talagrand, O. 1998. \icarus, 131, 302. \dodoi{10.1006/icar.1997.5874}


\bibitem[Forster et al.(2007)]{Forster2007}Forster, P., Ramaswamy, V., Artaxo,  P., et al. \ 2007.\  In: Climate Change 2007. Solomon, S., et al. eds. Cambridge University Press. \url{http://www.ipcc.ch/publications_and_data/ar4/wg1/en/ch2.html}


\bibitem[Fraedrich et al.(2005)]{Fraedrich2005} Fraedrich, K., Jansen, H., Kirk, E., Luksch, U., Lunkeit, F.\ 2005.\ The Planet Simulator: Towards a user friendly model.\ Meteo. Zeit.\ 14, 299. \dodoi{10.1127/0941-2948/2005/0043}

\bibitem[Fraedrich \& Lunkeit(2008)]{Fraedrich2008} Fraedrich, K., \& Lunkeit, F. 2008. TellA, 60, 921. \dodoi{10.1111/j.1600-0870.2008.00338.x}

\bibitem[Frierson(2007)]{Frierson2007} Frierson, D.~M. 2007. \jas., 64, 1959. \dodoi{10.1175/JAS3935.1}

\bibitem[Garcia \& Solomon(1983)]{Garcia1983} Garcia, R.~R., \& Solomon, S. 1983. \jgr, 88, 1379. \dodoi{10.1029/JC088iC02p01379}


\bibitem[Gray, Rumbold \& Shine (2009)]{Gray2009} Gray, L.~J., Rumbold, S.~T. and Shine, K.~P. 2009. \jas, 66, 2402. \dodoi{10.1175/2009JAS2866.1}

\bibitem[Green(1964)]{Green1964} Green, A.~E.~S. 1964. ApOpt, 3, 203. \dodoi{10.1364/AO.3.000203}

\bibitem[Hersbach et al.(2015)]{Hersbach2015} Hersbach, H., Peubey, C., Simmons, A., et al. 2015. QJRMS, 141, 2350. \dodoi{10.1002/qj.2528}

\bibitem[Ingersoll(1969)]{Ingersoll1969} Ingersoll, A.~P. 1969. \jas, 26, 1191. \dodoi{10.1175/1520-0469(1969)026<1191:TRGAHO>2.0.CO;2}

\bibitem[Kain(2004)]{Kain2004} Kain, J.~S. 2004. JApMe, 43, 170. \dodoi{10.1175/1520-0450(2004)043<0170:TKCPAU>2.0.CO;2}

\bibitem[Kasting and Donahue(1980)]{Kasting1980} Kasting, J.~F., Donahue, T.~M. 1980. \jgr, 85, 3255. \dodoi{10.1029/JC085iC06p03255}

\bibitem[Kasting and Donahue(1981)]{Kasting1981} Kasting, J.~F., Donahue, T.~M. 1981. Evolution of Oxygen and Ozone in Earths Atmosphere. NASA Conference Publication 2156, 149.

\bibitem[Kasting et al.(1984)]{Kasting1984} Kasting, J.~F., Pollack, J.~B., \& Ackerman, T.~P. 1984. \icarus 57, 335. \dodoi{10.1016/0019-1035(84)90122-2}

\bibitem[Kasting(1988)]{Kasting1988} Kasting, J.~F. 1988. \icarus, 74, 472. \dodoi{10.1016/0019-1035(88)90116-9}

\bibitem[Kasting et al.(1993)]{Kasting1993} Kasting, J.~F., Whitmire, D.~P., \& Reynolds, R.~T. 1993. \icarus, 101, 108. \dodoi{10.1006/icar.1993.1010}

\bibitem[Kasting et al.(2015)]{Kasting2015} Kasting, J.~F., Chen, H., \& Kopparapu, R.~K. 2015. \apj, 813, L3. \dodoi{10.1088/2041-8205/813/1/L3}

\bibitem[Katayama(1972)]{Katayama1972}Katayama, A., 1972. Tech. Report, No. 6 (Los Angeles, CA: UCLA Department of Meteorology)

\bibitem[Kopparapu et al.(2013)]{Kopparapu2013} Kopparapu, R.~K., Ramirez, R., Kasting, et al. 2013. \apj, 765, 131. \dodoi{10.1088/0004-637X/765/2/131}

\bibitem[Kuo(1965)]{Kuo1965} Kuo, H.~L. 1965. \jas, 22, 40. \dodoi{10.1175/1520-0469(1965)022<0040:OFAIOT>2.0.CO;2}

\bibitem[Kuo(1974)]{Kuo1974} Kuo, H.~L. 1974. \jas, 31, 1232. \dodoi{10.1175/1520-0469(1974)031<1232:FSOTPO>2.0.CO;2} 

\bibitem[Lacis and Hansen(1974)]{Lacis1974} Lacis, A.~A., Hansen, J. 1974. \jas, 31, 118. \dodoi{10.1175/1520-0469(1974)031<0118:APFTAO>2.0.CO;2}

\bibitem[Lane et al.(1973)]{Lane1973} Lane, A.~L., Barth, C.~A., Hord, C.~W., \& Stewart, A.~I. 1973.\icarus, 18, 102. \dodoi{10.1016/0019-1035(73)90175-9}

\bibitem[Laursen and Eliasen(1989)]{Laursen1989} Laursen, L., Eliasen, E. 1989. TellA, 41, 385. \dodoi{10.3402/tellusa.v41i5.11848}

\bibitem[Leconte et al.(2013)]{Leconte2013} Leconte, J., Forget, F., Charnay, B., Wordsworth, R., \& Pottier, A. 2013. \nat, 504, 268. \dodoi{10.1038/nature12827}

\bibitem[Linsenmeier et al.(2015)]{Linsenmeier2015} Linsenmeier, M., Pascale, S., \& Lucarini, V. 2015. P\&SS, 105, 43. \dodoi{10.1016/j.pss.2014.11.003}

\bibitem[Louis(1979)]{Louis1979} Louis, J.-F. 1979. BoLMe, 17, 187. \dodoi{10.1007/BF00117978}

\bibitem[Lucarini et al.(2010)]{Lucarini2010} Lucarini, V., Fraedrich, K., Lunkeit, F. 2010. ACPD, 10, 3699. \dodoi{10.5194/acp-10-9729-2010}

\bibitem[Lucarini and Ragone(2011)]{Lucarini2011} Lucarini, V., Ragone, F. 2011. RvGeo, 49, RG1001. \dodoi{10.1029/2009RG000323} 

\bibitem[Lucarini et al.(2013)]{Lucarini2013} Lucarini, V., Pascale, S., Boschi, R., Kirk, E., \& Iro, N. 2013. \ana, 334, 576. \dodoi{10.1002/asna.201311903}

\bibitem[Lunkeit et al.(2011)]{Lunkeit2011} Lunkeit, F., Fraedrich, K., Jansen, H., et al. 2011. Planet Simulator, Reference Manual. Technical Report, University of Hamburg. \url{https://www.mi.uni-hamburg.de/en/arbeitsgruppen/theoretische-meteorologie/modelle/sources/psusersguide.pdf}

\bibitem[Manabe \& M\"{o}ller(1961)]{Manabe1961}Manabe, S. \& M\"{o}ller, F. 1961. MWRv, 89, 503. \dodoi{10.1175/1520-0493(1961)089<0503:OTREAH>2.0.CO;2}

\bibitem[Manabe et al.(1965)]{Manabe1965} Manabe, S., Smagorinsky, J., \& Strickler, R.~F. 1965. MWRv, 93, 769. \dodoi{10.1175/1520-0493(1965)093<0769:SCOAGC>2.3.CO;2} 

\bibitem[Montmessin et al.(2011)]{Montmessin2011} Montmessin, F., Bertaux, J.-L., Lef{\`e}vre, F., et al. 2011. \icarus, 216, 82. \dodoi{10.1016/j.icarus.2011.08.010.}

\bibitem[Orszag(1970)]{Orszag1970} Orszag, S.~A. 1970. \jas, 27, 890. \dodoi{10.1175/1520-0469(1970)027<0890:TMFTCO>2.0.CO;2}  

\bibitem[Pascale et al.(2011)]{Pascale2011} Pascale, S., Gregory, J.~M., Ambaum, M.,  \& Tailleux, R. 2011. \cldy, 36, 1189. \dodoi{10.1007/s00382-009-0718-1} 

\bibitem[Poli et al.(2016)]{Poli2016} Poli, P., Hersbach, H., Dee, D. P., et al. 2016. \jcli, 29, 4083. \dodoi{10.1175/JCLI-D-15-0556.1} 

\bibitem[Popp et al.(2016)]{Popp2016} Popp, M., Schmidt, H., Marotzke, J. 2016.  \natco, 7, 10627. \dodoi{10.1038/ncomms10627}

\bibitem[Ragone et al.(2016)]{Ragone2016} Ragone, F., Lucarini, V., \& Lunkeit, F. 2016. \cldy, 46, 1459. \dodoi{10.1007/s00382-015-2657-3}

\bibitem[Ramirez and Kaltenegger(2014)]{Ramirez2014} Ramirez, R.~M., \& Kaltenegger, L. 2014. \apjl, 797, L25. \dodoi{10.1088/2041-8205/797/2/L25}

\bibitem[Ramirez and Kaltenegger(2016)]{Ramirez2016} Ramirez, R.~M., \& Kaltenegger, L. 2016. \apj, 823, 6. \dodoi{10.3847/0004-637X/823/1/6} 

\bibitem[Rodgers(1967)]{Rodgers1967}Rodgers, C. D., 1967. QJRMS, 93, 43. \dodoi{10.1002/qj.49709339504}

\bibitem[Roeckner et al.(1992)]{Roeckner1992} Roeckner, E., Arpe, K., Rockel, B., et al. 1992. Max-Planck-Institut f{\"u}r Meteorologie, Hamburg. Technical report 93. \url{https://www.mpimet.mpg.de/fileadmin/publikationen/Reports/MPI-Report_93.pdf}

\bibitem[Sasamori(1968)]{Sasamori1968} Sasamori, T. 1968. JApMe, 7, 721. \dodoi{10.1175/1520-0450(1968)007<0721:TRCCFA>2.0.CO;2} 

\bibitem[Slingo and Slingo(1991)]{Slingo1991} Slingo, A., Slingo, J.~M. 1991. \jgr, 96, 15. \dodoi{10.1029/91JD00930}

\bibitem[Stephens(1978)]{Stephens1978} Stephens, G.~L. 1978. \jas, 35, 2123. \dodoi{10.1175/1520-0469(1978)035<2123:RPIEWC>2.0.CO;2}

\bibitem[Stephens et al.(1984)]{Stephens1984} Stephens, G.~L., Ackerman, S., Smith, E.~A. 1984. \jas, 41, 687. \dodoi{10.1175/1520-0469(1984)041<0687:ASPRTI>2.0.CO;2}

\bibitem[Tian et al.(2009)]{Tian2009} Tian, W., Chipperfield, M.~P., L{\"u}, D. 2009. AdAtS, 26, 423. \dodoi{10.1007/s00376-009-0423-3}

\bibitem[Tiedtke(1988)]{Tiedtke1988} Tiedtke, M. \ 1988.\ In Physically-Based Modelling and Simulation of Climate and Climatic Change. Schlesinger M.E., ed. (Dordrecht: Springer), 375. \url{https://www.springer.com/la/book/9789027727886.}

\bibitem[Tiedtke(1989)]{Tiedtke1989} Tiedtke, M. 1989. MWRv, 117, 1779. \dodoi{10.1175/1520-0493(1989)117<1779:ACMFSF>2.0.CO;2.} 

\bibitem[Towe(1981)]{Towe1981} Towe, K.~M. 1981. PreR, 16, 1. \dodoi{10.1016/0301-9268(81)90002-4.}

\bibitem[Wilcox et al.(2012)]{Wilcox2012} Wilcox, L. J., Hoskins, B. J.\& Shine, K. P.   2012. QJRMS, 138, 561. \dodoi{10.1002/qj.951}

\bibitem[Wolf \& Toon(2015)]{Wolf2015} Wolf, E.~T., \& Toon, O.~B. 2015. \jgrd, 120, 5775. \dodoi{10.1002/2015JD023302} 

\bibitem[Wolf et al.(2017)]{Wolf2017} Wolf, E.~T., Shields, A.~L., Kopparapu, R.~K., Haqq-Misra, J., Toon, O.~B. 2017. \apj, 837, 107. \dodoi{10.3847/1538-4357/aa5ffc}

\bibitem[Wordsworth and Pierrehumbert(2013)]{Wordsworth2013} Wordsworth, R.~D., \& Pierrehumbert, R.~T. 2013. \apj, 778, 154. \dodoi{10.1088/0004-637X/778/2/154}

\bibitem[Yang et al.(2014)]{Yang2014} Yang, J., Bou{\'e}, G., Fabrycky, D.~C., \& Abbot, D.~S. 2014. \apjl, 787, L2. \dodoi{10.1088/2041-8205/787/1/L2}

\bibitem[Zhang \& MacFarlane(1995)]{Zhang1995} Zhang, G.~J., \& McFarlane N.~A. 1995. \ato, 33, 407. \dodoi{10.1080/07055900.1995.9649539}

\end{thebibliography}

\end{document}